\documentclass[useAMS,usenatbib]{mnras}

\usepackage{graphicx}	
\usepackage{subcaption}
\usepackage{floatrow}
\usepackage{amsmath}	
\usepackage{amssymb}	
\usepackage{multicol}        
\usepackage{bm}		
\usepackage{lipsum}
\usepackage{color}
\definecolor{blue}{RGB}{50, 80, 255}
\definecolor{red}{RGB}{255, 50, 50}

\newcommand{\dif}{\mathrm{d}}
\newcommand{\yr}{\,\mathrm{yr}}
\newcommand{\Myr}{\,\mathrm{Myr}}
\newcommand{\Gyr}{\,\mathrm{Gyr}}
\newcommand{\AU}{\,\mathrm{au}}
\newcommand{\ME}{M_{\oplus}}

\newcommand{\MJ}{M_{\rm J}}

\newcommand{\MSol}{M_{\odot}}
\newcommand{\RSol}{R_{\odot}}

\title[Secular chaos in WD planetary systems]{Secular chaos in white-dwarf planetary systems: origins of metal pollution and short-period planetary companions}

\author[O'Connor, Teyssandier \& Lai]
{Christopher E.\ O'Connor$^{1}$\thanks{E-mail: coconnor@astro.cornell.edu}, Jean Teyssandier$^{2}$, Dong Lai$^{1}$ \\
$^{1}$Cornell Center for Astrophysics and Planetary Science, Department of Astronomy, Cornell University, Ithaca, NY 14853, U.S.A. \\
$^{2}$Namur Institute for Complex Systems, Department of Mathematics, University of Namur, 61 Rue de Bruxelles, B-5000 Namur, Belgium
}

\begin{document}

\date{Accepted 2022 April 24. Received 2022 April 23; in original form 2021 November 15.}

\pagerange{\pageref{firstpage}--\pageref{lastpage}} \pubyear{2022}

\maketitle

\label{firstpage}

\begin{abstract}
    Secular oscillations in multi-planet systems can drive chaotic evolution of a small inner body through non-linear resonant perturbations.
    This ``secular chaos'' readily pushes the inner body to an extreme eccentricity, triggering tidal interactions or collision with the central star.
    We present a numerical study of secular chaos in systems with two planets and test particles using the ring-averaging method, with emphasis on the relationship between the planets' properties and the time-scale and efficiency of chaotic diffusion. 
    We find that secular chaos can excite extreme eccentricities on time-scales spanning several orders of magnitude in a given system. 
    We apply our results to the evolution of planetary systems around white dwarfs (WDs), specifically the tidal disruption and high-eccentricity migration of planetesimals and planets. 
    We find that secular chaos in a planetesimal belt driven by large ($\gtrsim 10 \ME$), distant ($\gtrsim 10 \AU$) planets can sustain metal accretion onto a WD over Gyr time-scales. 
    We constrain the total mass of planetesimals initially present within the chaotic zone by requiring that the predicted mass delivery rate to the Roche limit be consistent with the observed metal accretion rates of WDs with atmospheric pollution throughout the cooling sequence. 
    Based on the occurrence of long-period exoplanets and exo-asteroid belts, we conclude that secular chaos can be a significant (perhaps dominant) channel for polluting solitary WDs. 
    Secular chaos can also produce short-period planets and planetesimals around WDs in concert with various circularization mechanisms. 
    We discuss prospects for detecting exoplanets driving secular chaos around WDs using direct imaging and microlensing.
\end{abstract}

\begin{keywords}
    celestial mechanics -- planets and satellites: dynamical evolution and stability -- planetary systems -- white dwarfs
\end{keywords}

\section{Introduction} \label{s:intro}

Stars born with masses less than $\sim 8 \MSol$ are destined to become white dwarfs (WDs). 
This includes virtually all stars currently known to host planets or planet candidates. 
The survival and continued evolution of planetary systems during the final phases of the stellar life cycle is therefore of significant interest \citep[e.g.,][]{Veras2016}. 
The presence of remnant planetary systems around WDs has been inferred from the phenomenon of atmospheric metal pollution, where heavy elements are present in the otherwise-pure hydrogen or helium atmospheres of WDs. 
Pollution has been detected in 25 to 50 per cent of solitary WDs \citep{Zuckerman+2003,Zuckerman+2010,KGF2014} and in a similar fraction of wide binary systems containing a WD \citep{Zuckerman2014,Wilson+2019}. 
The source of the pollution is thought to be the tidal disruption and subsequent accretion of planetary bodies (either planetesimals or true planets) associated with the WD \citep[e.g.,][]{Jura2003}. 
This has been corroborated in recent years by observations of WDs with disintegrating objects currently in orbit \citep{Vanderburg+2015,Manser+2019,Gansicke+2019,Vanderbosch+2020,Farihi+2022}. 
Additionally, direct observations of intact planets orbiting WDs have been reported, following the discovery of the transiting planet candidate WD\,1856+534\,b \citep{Vanderburg+2020} and the microlensing planet candidate MOA-2010-BLG-477Lb \citep{Blackman+2021}.

Observational and theoretical studies of polluted WDs and their companions have provided insight into the physical properties, chemical composition, and dynamical evolution of extrasolar planets and small bodies \citep[e.g.,][]{Zuckerman+2007,Xu+2019b,TW2020,Bonsor+2020,OL2020}. 
Pollution probes mainly the cold outer regions of extrasolar planetary systems because objects orbiting closer than a few au during the host star's main-sequence phase are likely to be engulfed (and presumably destroyed) during post-main-sequence evolution \citep[e.g.,][]{VL2007,MV2012}. 
The delivery of planetary debris from these large distances is thought to be driven by massive, possibly unseen companions of the WD. 
The companions' gravitational perturbations excite the orbits of planetesimals to extreme eccentricities, causing them to be disrupted by tidal forces and subsequently accreted by the WD. 
Many variations of this scenario have been studied, such as planet--planet scattering triggered by AGB mass loss \citep{DS2002,MVV2014,Mustill+2018,VG2015,Veras+2016,Maldonado+2020a,Maldonado+2020b}, secular perturbations from surviving planets \citep[][]{BMW2011,DWS2012,FH2014,PML2017,Smallwood+2018}, and Lidov--Kozai oscillations induced by a distant stellar binary partner \citep*{HPZ2016,PM2017,SNZ2017}. 
Destabilization of giant planets' satellites has also been suggested (\citealt{Payne+2016,Payne+2017}; see also \citealt{Klein+2021}, \citealt{DDY2021}).

A successful model for the dynamical origins of WD pollution must satisfy several observational constraints. 
Firstly, the proposed mechanism must deliver polluting debris to the Roche limit around the WD (roughly $\sim \RSol$ for rocky rubble-piles) at a sufficient rate to account for the observed range of metal accretion rates among polluted WDs, roughly $\dot{M}_{Z} \sim 10^{6}$--$10^{12} \, {\rm g \, s^{-1}}$ \citep[e.g.,][]{KGF2014,Xu+2019b}. 
Secondly, the mechanism must be able to sustain the pollution over Gyr time-scales. 
The distribution of metal accretion rates is essentially independent of cooling age between $\sim 50 \Myr$ and $\sim 1 \Gyr$ \citep{Wyatt+2014,KGF2014,Xu+2019b}. 
Accretion rates appear to decrease for cooling ages greater than $\sim 1 \Gyr$, but the inferred degree of decline is sensitive to the assumptions made to calculate total accretion rates from measured chemical abundances \citep{HGK2018,Chen+2019,Xu+2019b,BX2022}.

We propose a new dynamical mechanism to drive WD pollution, which we have found to be capable of sustaining pollution over Gyr time-scales in a large fraction of remnant planetary systems. 
This is {\it secular chaos}, which arises from non-linear secular interactions between planetary orbits. 
The theory of secular chaos developed from studies of the long-term evolution and stability of the Solar System and of extrasolar planets \citep[e.g.][]{Laskar1997,Laskar2008,LW2011,LW2014,Laskar2017,VM2020,ML2021}. 
The simplest form of secular chaos occurs when two planets with moderately eccentric and inclined orbits perturb the motion of an inner test particle (a planetesimal or a smaller planet), leading to slow, diffusive growth of its eccentricity until tidal interactions with the central star become possible.

Secular chaos has been studied before as a possible avenue to produce highly eccentric orbits for exoplanets. 
\citet{WL2011} proposed that secular chaos in systems of three gas giants could produce hot Jupiters through high-eccentricity migration with tidal friction. 
Subsequent studies by \citet{Hamers+2017} and \citet*{TLV2019} found that chaotic secular evolution frequently results in tidal disruption of the would-be hot Jupiter, even when one accounts for short-range forces that tend to curtail eccentricity growth \citep*[e.g.,][]{LML2015}. 
The same studies found that extreme eccentricities could take up to a few Gyr to emerge, due to the diffusive nature of the chaos. 
These features lead us to ask to what extent secular chaos might also contribute to the pollution of WDs by planetesimals in remnant planetary systems.

In this paper, we demonstrate the favourable qualities of secular chaos as a dynamical mechanism for WD pollution. 
Among these qualities are (i) that secular chaos can deliver planetesimals steadily to the WD over Gyr time-scales; (ii) that it operates for planetary architectures that resemble the known population of long-period extrasolar planets; and (iii) that it may be distinguishable from other proposed pollution mechanisms by forthcoming observations. 
In Section \ref{s:theory}, we review the conditions that produce secular chaos in a system with two planets and an inner planetesimal. 
We also quantify the possible role of short-range forces in suppressing tidal disruption. 
In Section \ref{s:expts}, we describe a series of numerical experiments to characterize the relationship between the physical properties and orbital architecture of the original system and the rate of tidal disruption during secular chaos. 
In Section \ref{s:discuss}, we compare our results to observations of WD pollution and suggest applications in the dynamical evolution of remnant planetary systems more broadly. 
In Section \ref{s:conclude}, we summarize our main conclusions.

\section{Secular Dynamics, Chaos, and Tidal Disruption} \label{s:theory}

Chaos emerges in systems of coupled oscillators from the confluence of resonance and non-linearity \citep[e.g.,][]{Chirikov1979}. 
We consider a test mass surrounded by two outer planets. 
Secular resonances occur when the free precession frequency of the test mass matches the natural frequency of a secular mode of the outer planets, leading to a large forced eccentricity or inclination. 
Non-linear interactions become important when the test mass' eccentricity or inclination is large. 
These introduce coupling between secular modes, leading to chaotic diffusion of the test mass' eccentricity.

\subsection{Linear secular theory and onset of chaos}

Consider a planetary system with a central mass $M_{*}$, two planets (labelled $j = \{ 1, 2 \}$), and an inner test particle (or ``planetesimal,'' labelled `p'). 
Let the planets' masses be $m_{j} \ll M_{*}$, and let the usual Keplerian orbital elements of each body be $(a_{j}, e_{j}, I_{j}, \varpi_{j}, \Omega_{j})$. 
For each orbit, we define the complex-valued eccentricity and inclination as
\begin{subequations}
\begin{align}
    \tilde{e}_{j} &\equiv e_{j} \exp( \iota\varpi_{j} ), \\
    \tilde{I}_{j} &\equiv I_{j} \exp( \iota\Omega_{j} ).
\end{align}
\end{subequations}
The secular planet--planet interactions in this system can be expressed in terms of the quadrupole and octupole precession frequencies:
\begin{align}
    \omega_{jk} &= \frac{G m_{j} m_{k} a_{<}}{4 L_{j} a_{>}^{2}} b_{3/2}^{(1)}(\alpha), \label{eq:om_jk} \\
    \nu_{jk} &= \frac{G m_{j} m_{k} a_{<}}{4 L_{j} a_{>}^{2}} b_{3/2}^{(2)}(\alpha), \label{eq:nu_jk}
\end{align}
where $a_{<} = \min(a_{j},a_{k})$, $a_{>} = \max(a_{j},a_{k})$, $\alpha = a_{<}/a_{>}$, and $L_{j} \simeq m_{j} (G M_{*} a_{j})^{1/2}$. 
The functions $b_{3/2}^{(1)}$ and $b_{3/2}^{(2)}$ are Laplace coefficients, given by
\begin{equation}
    b_{3/2}^{(q)}(\alpha) = \frac{1}{\pi} \int_{0}^{2 \pi} \frac{\cos(qx) \, \dif x}{(1 + \alpha^{2} - 2 \alpha \cos{x})^{3/2}}.
\end{equation}
For small eccentricities and mutual inclinations, the secular equations of motion are as follows \citep[e.g.,][]{MD1999,PL2018}:
\begin{subequations} \label{eq:secular_matrices}
\begin{align}
    \frac{\dif}{\dif t} \begin{bmatrix}
        \tilde{e}_{\rm p} \\ \tilde{e}_{1} \\ \tilde{e}_{2} 
    \end{bmatrix} &= \iota
    \begin{bmatrix}
        \omega_{\rm p} & -\nu_{\rm p1} & -\nu_{\rm p2} \\
        0 & \omega_{12} & -\nu_{12} \\
        0 & -\nu_{21} & \omega_{21} 
    \end{bmatrix}
    \begin{bmatrix}
        \tilde{e}_{\rm p} \\ \tilde{e}_{1} \\ \tilde{e}_{2} 
    \end{bmatrix}, \label{eq:matrix_ecc} \\
    \frac{\dif}{\dif t} \begin{bmatrix}
        \tilde{I}_{\rm p} \\ \tilde{I}_{1} \\ \tilde{I}_{2} 
    \end{bmatrix} &= \iota
    \begin{bmatrix}
        -\omega_{\rm p} & \omega_{\rm p1} & \omega_{\rm p2} \\
        0 & -\omega_{12} & \omega_{12} \\
        0 & \omega_{21} & -\omega_{21} 
    \end{bmatrix}
    \begin{bmatrix}
        \tilde{I}_{\rm p} \\ \tilde{I}_{1} \\ \tilde{I}_{2}
    \end{bmatrix}, \label{eq:matrix_incl}
\end{align}
\end{subequations}
where $\omega_{\rm p} = \omega_{\rm p1} + \omega_{\rm p2}$. 
The linear eigenmodes and eigenvalues are obtained by diagonalizing the coefficient matrices in Eqs.\ (\ref{eq:secular_matrices}). 
Solving Eq.\ (\ref{eq:matrix_ecc}) yields three normal modes for the system's eccentricity oscillations. 
Two of these modes, labelled I and II, describe the secular oscillations of the planets:
\begin{equation} \label{eq:ecc_modes}
    \tilde{e}_{j}(t) = \sum_{\alpha={\rm I,II}} c_{\alpha} v_{\alpha,j} \exp(\iota \lambda_{\alpha} t),
\end{equation}
where $j = \{1,2\}$, $\boldsymbol{v}_{\alpha}$ is the eigenvector of the coefficient matrix in Eq.\ (\ref{eq:matrix_ecc}) with eigenvalue $\lambda_{\alpha}$, and
\begin{equation} \label{eq:lambda_pm}
    \lambda_{\rm I,II} = \frac{1}{2} \left\{ (\omega_{12} + \omega_{21}) \pm \left[ (\omega_{12} - \omega_{21})^{2} + 4 \nu_{12} \nu_{21} \right]^{1/2} \right\}.
\end{equation}
The complex-valued secular mode amplitudes $c_{\alpha}$ are given by linear combinations of the initial conditions $\Tilde{e}_{j,0}$. 
In practice, $c_{\rm I}$ is dominated by $\Tilde{e}_{1,0}$ and $c_{\rm II}$ by $\Tilde{e}_{2,0}$. 
Meanwhile, the third mode, with eigenvalue $\omega_{\rm p}$, represents the free apsidal precession of the test mass driven by the two outer planets. 
The formal solution for $\tilde{e}_{\rm p}(t)$ can be written as
\begin{align}
    \tilde{e}_{\rm p}(t) &= A \exp(\iota \omega_{\rm p} t) + \frac{B_{\rm I}}{\omega_{\rm p} - \lambda_{\rm I}} \exp(\iota \lambda_{\rm I} t) \nonumber \\
    & \hspace{0.5cm} + \frac{B_{\rm II}}{\omega_{\rm p} - \lambda_{\rm II}} \exp(\iota \lambda_{\rm II} t), \label{eq:e_secular_soln}
\end{align}
where $A$ is a complex-valued constant describing the amplitude of the free eccentricity mode and $B_{\rm I}$ and $B_{\rm II}$ are linear combinations of the octupole frequencies $\nu_{{\rm p}j}$ and the secular mode amplitudes $c_{\rm I,II}$.

The eigenmodes of Eq.\ (\ref{eq:matrix_incl}) describe the nodal precession of each body. 
One mode has eigenvalue zero and can be eliminated without loss of generality by measuring all orbital inclinations with respect to the system's invariable plane. 
The remaining modes have eigenvalues $\lambda_{0} = -(\lambda_{\rm I} + \lambda_{\rm II}) = -(\omega_{12} + \omega_{21})$ and $-\omega_{\rm p}$; the former describes the precession of the mutually inclined outer planets, while the latter describes the free nodal precession of the test particle due to the outer planets' combined perturbation. 
Analogously to Eq.\ (\ref{eq:e_secular_soln}), the inclination of the test mass evolves according to
\begin{equation}
    \tilde{I}_{\rm p}(t) = C \exp(-\iota \omega_{\rm p} t) + \frac{D}{\omega_{\rm p} - |\lambda_{0}|} \exp(-\iota |\lambda_{0}| t),
\end{equation}
where $C$ and $D$ are constants analogous to $A$ and $B_{\rm I,II}$. 
When $\omega_{\rm p}$ is close to one of the eigenfrequencies of the outer planets ($\lambda_{\rm I,II}$ or $|\lambda_{0}|$), the planetesimal is said to be in a linear secular resonance with the outer planets. 
Its forced eccentricity or inclination can grow to a large value, at which point secular interactions can become non-linear.

\begin{figure}
    \centering
    \includegraphics[width=\columnwidth]{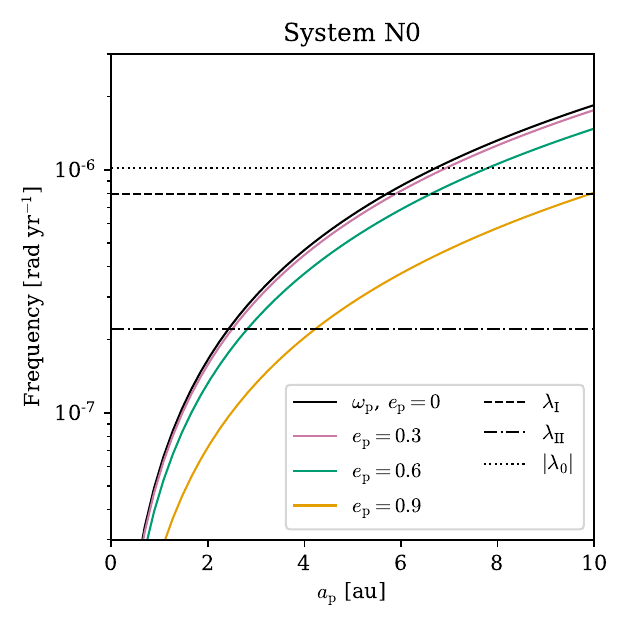}
    \caption{Characteristic secular frequencies for the problem of a planetesimal forced by two outer planets. 
    The solid curves show the free precession rate $\omega_{\rm p}$ of a test particle with semi-major axis $a_{\rm p}$ (Eq.\ \ref{eq:nonlinear_gamma}) for various initial eccentricities of the planetesimal. 
    The horizontal lines mark the Laplace--Lagrange eigenfrequencies $\lambda_{\rm I,II}$ and $|\lambda_{0}|$ of the planets. 
    Secular resonances occur where $\omega_{\rm p} \approx \lambda_{i}$. 
    For small initial eccentricities $e_{\rm p}$, secular chaos is expected to occur for $a_{\rm p} \approx 4$--$8 \AU$ to the overlap of the resonances $\omega_{\rm p} \approx \lambda_{\rm I}$ and $\omega_{\rm p} \approx |\lambda_{0}|$.}
    \label{fig:secular_freqs}
\end{figure}

In the non-linear regime, there is no single expression for a planetsimals's free precession rate. 
However, by examining the non-linear secular equations \citep[e.g.,][]{LML2015}, one can make an `educated guess' for the instantaneous precession rate of an orbit with eccentricity $e_{\rm p}$ and inclination $I_{\rm p}$:
\begin{equation} \label{eq:nonlinear_gamma}
    \omega_{\rm p}^{\rm (NL)} \sim \omega_{\rm p} \left( 1-e_{\rm p}^{2} \right)^{1/2} \cos{I_{\rm p}}
\end{equation}
Chaos can arise in regions where non-linear secular resonances overlap \citep[e.g.,][]{LW2011}. 
Although Eq.\ (\ref{eq:nonlinear_gamma}) is not a rigorous formula, we can use it to estimate the location of the chaotic zone given a configuration of planets \citep[e.g.,][]{TLV2019}. 
As an example, consider a system where the planets have masses $(m_{1},m_{2}) = (30,20) \ME$ and semi-major axes $(a_{1},a_{2}) = (20,43) \AU$, with small eccentricities and nearly coplanar orbits. 
The three eigenfrequencies of this system are
\begin{equation*}
    \lambda_{\rm I} \simeq \frac{2 \pi}{5.4 \Myr}, \  \lambda_{\rm II} \simeq \frac{2 \pi}{25 \Myr}, \ |\lambda_{0}| \simeq \frac{2 \pi}{4.4 \Myr}.
\end{equation*}
To estimate the location at which a test particle will develop secular chaos, we calculate the non-linear precession rate as a function of the semi-major axis $a_{\rm p}$ using Eq.\ (\ref{eq:nonlinear_gamma}) for different eccentricities $e_{\rm p}$ and compare the result to the eigenfrequencies. 
For simplicity, we assume in this example that the particle is coplanar with the perturbers, so as to focus on the nonlinear effects in eccentricity only. 
Figure \ref{fig:secular_freqs} displays the result of this exercise. 
We see that, depending on the test particle's eccentricity, non-linear secular resonances occur in a range of semi-major axes. 
Note in particular the close proximity of the linear apsidal precession resonance $\omega_{\rm p} = \lambda_{\rm I}$ and nodal precession resonance $\omega_{\rm p} = |\lambda_{0}|$ (where the solid curves intersect the horizontal dashed and dotted lines, respectively). 
This suggests that the non-linear resonances can overlap and drive secular chaos of the test particle. 
We therefore expect chaos to emerge when the test particle is placed somewhat interior to the linear nodal precession resonance, say in the range $\sim 4$--$8 \AU$ for the system in question. 
We verify this prediction with our numerical experiments in Section \ref{s:expts}.

\subsubsection{Chaotic diffusion time-scale}

One aspect of secular chaos we wish to investigate is the characteristic time-scale of a test particle's chaotic orbital diffusion (denoted $t_{\rm diff}$). 
This is relevant to our application of secular chaos to WD pollution because it relates to the rate at which planetesimals undergo tidal disruption as a function of a system's cooling age. 
While there is not a rigorous determination of $t_{\rm diff}$, we can estimate it with dimensional analysis.

The dynamics of a test particle near an apsidal precession resonance is driven by the octupole-order perturbation from the outer planets, for instance the mode dominated by the eccentricity of planet 1. 
Per Eq.\ (\ref{eq:secular_matrices}), we would expect the chaotic diffusion rate (defined as $1 / t_{\rm diff}$) to scale as
\begin{equation} \label{eq:tdiff}
    \frac{1}{t_{\rm diff}} \sim \nu_{\rm p1} e_{1} \sim \left( \frac{\mathcal{G} M_{*}}{a_{\rm p}^{3}} \right)^{1/2} \frac{m_{1}}{M_{*}} \left( \frac{a_{\rm p}}{a_{1}} \right)^{4} e_{1}.
\end{equation}
One goal of our numerical experiments (Section \ref{s:expts}) is to test the validity of this scaling relation. 
A possible shortcoming is that the true diffusion rate may also depend on a number of other dimensionless quantities, such as the mass ratio $m_{1}/m_{2}$, the semi-major axis ratio $a_{1}/a_{2}$, and the initial apsidal and nodal angles $\varpi_{j}$ and $\Omega_{j}$. 
Secular chaos emerges only close to the system's linear secular resonances, and thus Eq.\ (\ref{eq:tdiff}) relates the dynamics of two different systems only if the resonance condition $\omega_{\rm p} \approx \lambda_{\rm I} \approx |\lambda_{0}|$ is satisfied in both cases.

The linear secular theory is scale-free with respect to the masses and semi-major axes of the planets. 
Specifically, all secular frequencies depend on the quantity $m_{j} / a_{j}^{3/2}$ and various ratios such as $a_{\rm p} / a_{j}$. 
Thus, the secular dynamics of two systems are similar (for the same initial conditions in terms of eccentricities, inclinations, and phase angles) if their properties are related by a transformation of the form
\begin{align} \label{eq:secular_rescaling}
    m'_{j} = F m_{j}, \ \
    a'_{j} = F^{2/3} a_{j}, \ \ 
    a'_{\rm p} = F^{2/3} a_{\rm p}
\end{align}
with $F > 0$. 
This is not necessarily true in the non-linear, chaotic regime; it is certainly not true when integrating the motion numerically with finite precision, since numerical noise will lead to exponential separation of solutions. 
However, our numerical experiments in Section \ref{s:expts} show that a version of this scale invariance holds for ensembles of test particle--planet systems.

\subsection{Tidal disruption and short-range forces} \label{s:theory:SRFs}

Consider a planetesimal of mass $m_{\rm p}$ and radius $R_{\rm p}$, bound primarily by self-gravity. 
The critical distance from the central star (mass $M_{*}$) for tidal disruption (a.k.a.\ the Roche limit) is
\begin{equation}
    r_{\rm dis} = \eta R_{\rm p} \left( \frac{M_{*}}{m_{\rm p}} \right)^{1/3},
\end{equation}
where $\eta$ is a numerical factor determined by the planetesimal's composition, structure, and rotation. 
For a strengthless body such as a rubble-pile asteroid or a planet in hydrostatic equilibrium, $\eta$ is between $1$ and $3$ \citep[e.g.,][]{Davidsson1999,GRL2011}. 
On the other hand, for a solid body with high internal strength, $\eta$ can be smaller. 
We will assume a fiducial WD with $M_{*} = 0.6 \MSol$ and a rocky, strengthless planetesimal with a density of $\rho_{\rm p} = 3 \, {\rm g \, cm^{-3}}$. 
This gives $r_{\rm dis} \approx 0.66$--$2.0 \RSol$ for $1 < \eta < 3$, which agrees with the observed radii of dusty accretion disks around polluted WDs \citep*[e.g.,][]{FJZ2009}.

In order to reach the Roche limit at pericentre from a semi-major axis $a$, a planetesimal orbiting a WD must have eccentricity greater than a critical value $e_{\rm dis}$ given by
\begin{align} \label{eq:edis_def}
    1 - e_{\rm dis} &= \frac{\eta R_{\rm p}}{a_{\rm p}} \left( \frac{M_{*}}{m_{\rm p}} \right)^{1/3}, \nonumber \\
    &= 1.2 \times 10^{-3} \left( \frac{\eta}{2} \right) \left( \frac{a_{\rm p}}{5 \AU} \right)^{-1} \nonumber \\
    & \hspace{1.5cm} \times \left( \frac{M_{*}}{0.6 \MSol} \right)^{1/3} \left( \frac{\rho_{\rm p}}{3 \, {\rm g \, cm^{-3}}} \right)^{-1/3}.
\end{align}
In order to pollute a WD with planetesimals via secular chaos, planetesimals must be excited to $e_{\rm p} \geq e_{\rm dis}$ within the star's cooling age.

When the planetesimal approaches a small pericentre distance, short-range forces (SRFs) introduce additional perturbations that can suppress eccentricity excitation. 
In general, SRFs impose an upper limit on the eccentricity that a test particle can achieve through secular forcing \citep*[e.g.,][]{FT2007,LML2015,PML2017}. 
Thus, they can, in principle, act as a barrier to WD pollution by preventing planetesimals from reaching the Roche limit. 
An important SRF arises from the general relativistic (GR) correction of the stellar gravitational potential, which induces free apsidal precession at a rate
\begin{equation} \label{eq:def_om_GR}
    \omega_{\rm GR} = \frac{3 G M_{*}}{c^{2} a_{\rm p}} \frac{n_{\rm p}}{1 - e_{\rm p}^{2}},
\end{equation}
where $n_{\rm p} = (G M_{*} / a_{\rm p}^{3})^{1/2}$ is the particle's mean motion. 
The maximal eccentricity achievable by the test particle through secular forcing (or ``limiting eccentricity'' $e_{\rm lim}$) is determined by the competition between $\omega_{\rm GR}$ and the characteristic secular forcing rate given by \citep{LML2015}
\begin{equation} \label{eq:def_om_LK}
    \omega_{\rm LK} = \frac{n_{\rm p}}{(1-e_{\rm p}^{2})^{1/2}} \frac{a_{\rm p}^{3}}{M_{*}} \sum_{j} \frac{m_{j}}{a_{j}^{3}} \frac{1}{(1-e_{j}^{2})^{3/2}}.
\end{equation}
Setting $\omega_{\rm GR} \sim \omega_{\rm LK}$ with $e_{\rm p} = e_{\rm lim} \simeq 1$, we find
\begin{equation} \label{eq:elim_eps1PN}
    1 - e_{\rm lim} \sim \varepsilon_{\rm GR}^{2},
\end{equation}
where
\begin{equation} \label{eq:eps1PN_def}
    \varepsilon_{\rm GR} \equiv \frac{3 G M_{*}^{2}}{c^{2} a_{\rm p}^{4}} \left[ \sum_{j} \frac{m_{j}}{a_{j}^{3}} \frac{1}{(1-e_{j}^{2})^{3/2}} \right]^{-1}
\end{equation}
is a dimensionless quantity describing the magnitude of the GR correction relative to secular forcing. 
An accurate numerical coefficient for Eq.\ (\ref{eq:elim_eps1PN}) can be obtained analytically for specific forms of secular forcing, such as the octupole-order Lidov--Kozai effect \citep{LML2015} and secular driving inside mean-motion resonances \citep{PML2017}. 
However, for secular chaos, the coefficient may depend on the dimensionless orbital elements of the outer planets (e.g., eccentricities, $a_{1}/a_{2}$).

Additional SRFs can arise from the tidal and rotational distortion of the planetesimal's equilibrium figure. 
Comparing the precession rates associated with these forces (denoted $\omega_{\rm tide}$ and $\omega_{\rm rot}$ respectively) to the GR precession rate (see Eq.\ 57 of \citealt{LML2015}), we find
\begin{align}
    \frac{\omega_{\rm tide}}{\omega_{\rm GR}} &\sim k_{\rm 2p} \frac{c^{2} R_{\rm p}^{2}}{G \rho_{\rm p} q_{\rm p}^{4}} \nonumber \\
    &\simeq 10^{-4} k_{\rm 2p} \left( \frac{R_{\rm p}}{10 \, {\rm km}} \right)^{2} \left( \frac{\rho_{\rm p}}{3 \, {\rm g \, cm^{-3}}} \right)^{-1} \left( \frac{q_{\rm p}}{\RSol} \right)^{-4}, \\
    \frac{\omega_{\rm rot}}{\omega_{\rm GR}} &\sim k_{\rm q,p} \frac{c^{2} R_{\rm p}^{2}}{G \rho_{\rm p} (q_{\rm p}^{5} a_{\rm p}^{3})^{1/2}} \nonumber \\
    &\simeq 10^{-7} k_{\rm q,p} \left( \frac{R_{\rm p}}{10 \, {\rm km}} \right)^{2} \left( \frac{\rho_{\rm p}}{3 \, {\rm g \, cm^{-3}}} \right)^{-1} \nonumber \\
    & \hspace{1.5cm} \times \left( \frac{q_{\rm p}}{\RSol} \right)^{-5/2} \left( \frac{a_{\rm p}}{1 \AU} \right)^{-3/2},
\end{align}
where $k_{\rm 2p}$ and $k_{\rm q,p}$ are the tidal Love number and rotational distortion coefficient of the planetesimal and where $q_{\rm p} = a_{\rm p} (1 - e_{\rm p})$ is the pericentre distance. 
Thus, $\omega_{\rm tide}$ and $\omega_{\rm rot}$ are negligible compared to $\omega_{\rm GR}$ for planetesimals. 
Perturbations arising from the tidal and rotational distortion of the WD can be treated in a similar manner and are also negligible. 
Accordingly, our numerical experiments in Section \ref{s:expts} will include the GR correction but exclude the tidal and rotational SRFs.

\section{Numerical Experiments} \label{s:expts}

In this section, we describe our numerical experiments with systems of a test particle and two outer planets, with the primary goal of characterizing the relationship between the time-scale of secular chaos and the system's dynamical architecture. 
It is also of interest to know the fraction of simulated test particles that experience chaotic diffusion, since this quantity informs the translation of theoretical results to observations.

The behaviour of a chaotic dynamical system is sensitive to its initial conditions and a set of characteristic quantities. 
We set out to characterize the dependence of the chaotic-diffusion time-scale $t_{\rm diff}$ on three main quantities:
\begin{itemize}
    \item The planets' masses $m_{1}$ and $m_{2}$, with the ratio $m_{2}/m_{1}$ fixed.
    \item The initial eccentricity $e_{1}$ of the inner perturber, with all other perturber initial conditions ($e_{2}$ and all angles $I_{j}$, $\varpi_{j}$, $\Omega_{j}$) fixed.
    \item The ratio of the particle's and inner planet's semi-major axes $a_{\rm p}/a_{1}$, with $a_{1}$ and $a_{2}$ fixed.
\end{itemize}
This approach is motivated by Eq.\ (\ref{eq:tdiff}), which predicts the scaling of the chaotic diffusion rate with various quantities based on the linear secular theory. 
Each quantity acts as a control parameter for the particle's secular evolution in a different way. 
Re-scaling the outer planets' masses changes only the period of secular forcing, which sets the overall time-scale of chaotic diffusion. 
Adjusting the spacing between the test particle and the planets affects its forced eccentricity and inclination. 
Finally, changing a planet's initial eccentricity changes the secular mode amplitudes $c_{\rm I,II}$, which likewise affects the particle's forced eccentricity. 
Based on Fig.\ \ref{fig:secular_freqs}, adjusting the amplitude of eccentricity mode I has the most significant effect on the chaotic evolution of a test particle; this is why we vary the initial value of $e_{1}$, which is approximately proportional to $|c_{\rm I}|$ \citep[e.g.,][]{PL2019}. 
The initial values of $e_{2}$, $\varpi_{1}$, and $\varpi_{2}$ also affect $c_{\rm I}$ but are of secondary importance. 
Because chaos is driven by the non-linear coupling of eccentricity mode I with inclination mode 0, adjusting the planets' mutual inclination would have a similar effect as adjusting $e_{1}$.

Before describing our experiments, we must choose a suitable definition for the diffusion time-scale $t_{\rm diff}$. 
The theory of chaotic dynamical systems states that nearby trajectories in phase space diverge exponentially on the system's Lyapunov time-scale, which can be calculated from numerical simulations. 
Therefore, from a purely theoretical standpoint the Lyapunov time is the most rigorous choice for $t_{\rm diff}$. 
In our case, however, we are interested in the time-scale on which diffusion leads to tidal disruption on a highly eccentric orbit. 
Therefore, to estimate $t_{\rm diff}$, we record the time $t_{\rm dis}$ in each simulation at which the test particle's eccentricity reaches the threshold value for tidal disruption (Eq.\ \ref{eq:edis_def}). 
The value of $t_{\rm dis}$ can vary by orders of magnitude due to small changes in the initial conditions. 
We therefore should think of our numerical experiments as sampling the underlying distribution of a random variable $t_{\rm dis}$; the properties of this distribution presumably depend on the planetary system's configuration and the distribution of initial conditions. 
We will estimate $t_{\rm diff}$ as the median value of $t_{\rm dis}$ from numerical integrations.

\subsection{Methods and initial conditions} \label{s:expts:methods}

We base our numerical methods on those of \citet*{TLV2019}, who used the ring-averaging method to study high-eccentricity migration of giant planets through secular chaos. 
The ring-averaging method treats secular planet--planet interactions by `smearing' the mass of each body around its Keplerian orbit and computing the torques exchanged by each pair of the resulting elliptical rings. 
This technique combines a relatively low computational cost with a high degree of accuracy for arbitrary eccentricities and mutual inclinations, provided that there be no close encounters and that no bodies be in mean-motion resonance. 
We carry out our numerical experiments using the publicly available code {\sc rings},\footnote{\url{ https://github.com/farr/Rings}} written by Dr.~Will~M.~Farr based on the algorithm of \citet*{TTK2009}. 
We have modified the code to include the GR correction following \citet{TLV2019}.

Unless stated otherwise, our simulations were carried out as follows. 
Each experiment consisted of 1,000 trials. 
Each trial had a simulated duration of up to $10.0 \Gyr$ but was terminated early if the test particle reached an eccentricity of $e_{\rm dis}=0.999$. 
As we noted in Section \ref{s:theory:SRFs}, a fiducial planetesimal would be disrupted with an eccentricity $0.995 \lesssim e_{\rm dis} \lesssim 0.9995$ for initial semi-major axes $1 \AU \lesssim a_{\rm p} \lesssim 10 \AU$. 
We do not expect the GR correction to significantly influence the dynamics in our main experiments because the limiting eccentricity significantly exceeds the threshold for the architectures we consider, i.e.\ $1 - e_{\rm lim} \ll 10^{-3}$. 
We ignore the possibility of collisions between planetesimals throughout this work.

The initial conditions for each trial were generated as follows. In each experiment, the planets' masses and initial orbits were the same in all trials. 
The test particle's initial eccentricity and inclination were drawn from truncated Rayleigh distributions meant to approximate the Solar System's main asteroid belt \citep[e.g.,][]{RN2020}. 
The r.m.s.\ initial eccentricity and inclination were $0.12$ and $12^{\circ}$, respectively, and the distributions were truncated for $e_{\rm p} > 0.4$ and $I_{\rm p} > 40^{\circ}$. 
The longitudes of the periastron and the ascending node were both drawn from uniform distributions.

\subsubsection{Preliminary examples} \label{s:expts:prelim}

\begin{table}[]
    \centering
    \begin{tabular}{ccl}
        \hline
        Quantity & Unit & N0 Values \\ \hline
        $M_{*}$ & $\MSol$ & $0.6$ \\
        $m_{1,2}$ & $\ME$ & $30$, $20$ \\
        $a_{1,2}$ & au & $20.0$, $43.0$ \\
        $e_{1,2}$ & -- & $0.15$, $0.05$ \\
        $I_{1,2}$ & deg & $6.5$, $6.5$ \\ \hline
        Quantity & Unit & Distribution \\ \hline
        $e_{\rm p}$ & -- & $\mathcal{R}(0.12)$, $\max(e_{\rm p}) = 0.4$ \\
        $I_{\rm p}$ & deg & $\mathcal{R}(12)$, $\max(I_{\rm p}) = 40^{\circ}$ \\
        $\varpi_{\rm p}$ & rad & $\mathcal{U}(0,2\pi)$ \\
        $\Omega_{\rm p}$ & rad & $\mathcal{U}(0,2\pi)$
        \\ \hline
    \end{tabular}
    \caption{{\it Upper:} Initial conditions of the fiducial two-planet system N0. 
    {\it Lower:} Distributions of test-particle initial conditions used in all numerical experiments, unless otherwise specified. 
    The distributions of $a_{\rm p}$ are listed in Table \ref{tab:tpchaos_ICs}. 
    $\mathcal{U}(a,b)$ signifies a uniform distribution on the interval $[a,b)$. 
    $\mathcal{R}(x)$ denotes a Rayleigh distribution with scale parameter $x$.}
    \label{tab:N0_params}
\end{table}

Here we provide three illustrative examples of secular test-particle trajectories, called `Ex-1', `Ex-2', and `Ex-3'. 
In all cases we use a fiducial outer-planet architecture called N0, with parameters listed\footnote{In addition to the orbital elements listed in Table \ref{tab:N0_params}, we choose initial apsidal angles $\varpi_{1} = 93 \fdg 2$, $\varpi_{2} = 266 \fdg 9$. 
This choice minimizes the amplitude of eccentricity mode I ($|c_{\rm I}|$, see Eq.\ \ref{eq:ecc_modes}) for fixed initial values of $e_{1}$ and $e_{2}$ because $\varpi_{1} - \varpi_{2} \approx 180^{\circ}$. 
We will later show that mode I is mainly responsible for driving chaos in our simulations while mode II is not. 
However, the particular values of these angles have a small effect on the emergence of secular chaos. 
The nodal angles always satisfy $\Omega_{1} - \Omega_{2} = 180^{\circ}$.}
in Table \ref{tab:N0_params}.

\begin{figure}
    \centering
    \includegraphics[width=\columnwidth]{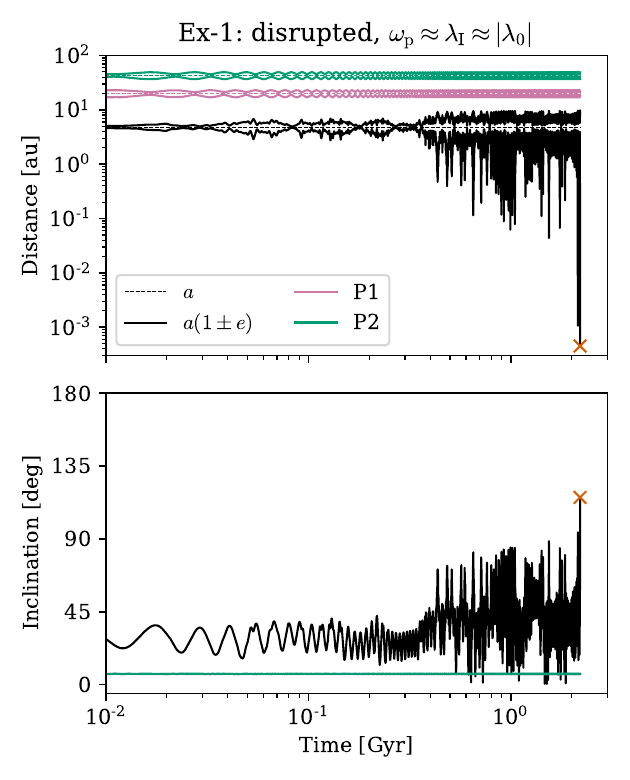}
    \caption{Secular evolution of the planetesimal (black curves) and planets (purple and green) in run Ex-1. 
    The particle's initial state is $a_{\rm p} = 4.8230 \AU$, $e_{\rm p} = 0.1557$, $I_{\rm p} = 17\fdg6780$, $\varpi_{\rm p} = 98\fdg3591$, $\Omega_{\rm p} = 348\fdg0021$. 
    The integration was halted at $t = 2.197 \Gyr$; the particle's terminal state is marked with a red cross. 
    {\it Upper:} Semi-major axes $a$ (dashed) and pericentre and apocentre distances $a (1 \pm e)$ (solid). 
    {\it Lower:} Inclinations relative to the invariable plane.}
    \label{fig:history_Ex1}
\end{figure}

Run Ex-1, shown in Figure \ref{fig:history_Ex1}, demonstrates a chaotic trajectory leading to tidal disruption of the planetesimal. 
The particle is initialized in a range of parameter space where the secular resonances $\omega_{\rm p} \approx \lambda_{\rm I}$ and $\omega_{\rm p} \approx |\lambda_{0}|$ overlap and chaos is expected. 
Initially, the trajectory has a quasiperiodic character, with the eccentricity and inclination undergoing small oscillations. 
Around $t = 400 \Myr$, however, the trajectory abruptly `breaks free' and exhibits chaotic diffusion, with sudden spikes in the eccentricity and inclination. 
Shortly before $t = 2.2 \Gyr$ an extreme eccentricity spike takes it to the threshold $e = 0.9999$ and the run is halted; we note that the particle's terminal orbit is retrograde, with an inclination $I = 115\fdg6$.

\begin{figure}
    \centering
    \includegraphics[width=\columnwidth]{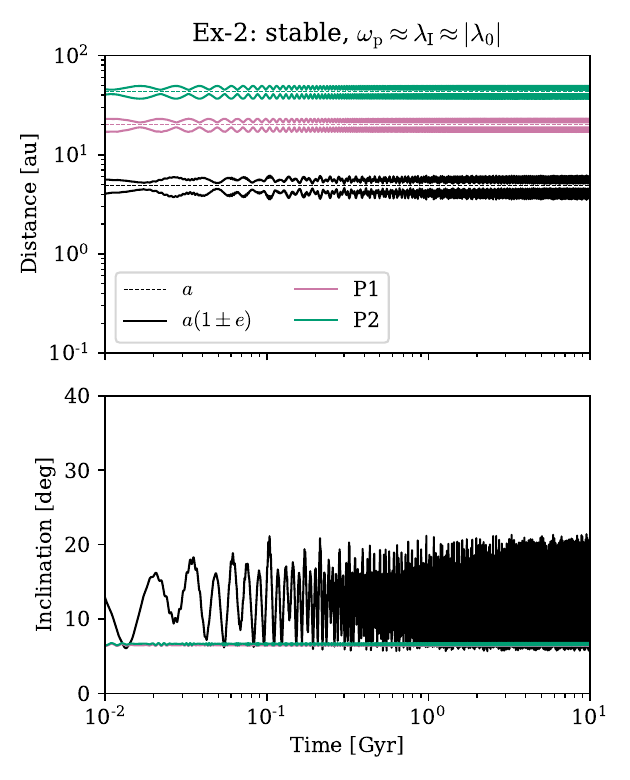}
    \caption{Same as Fig.\ \ref{fig:history_Ex1}, but for run Ex-2. 
    The particle's initial state is $a_{\rm p} = 4.8409 \AU$, $e_{\rm p} = 0.1510$, $I_{\rm p} = 7\fdg3370$, $\varpi_{\rm p} = 6\fdg7644$, $\Omega_{\rm p} = 196\fdg3190$. 
    The integration lasted the full $10.0 \Gyr$; the particle reaches a maximum eccentricity of $0.2744$.}
    \label{fig:history_Ex2}
\end{figure}

Run Ex-2 (Fig.\ \ref{fig:history_Ex2}) shows a trajectory within the chaotic zone that remains quasiperiodic over $10 \Gyr$ of secular evolution. 
The eccentricity and inclination never exceed $0.28$ and $22^{\circ}$ respectively. 
This example shows that a fraction of orbits in the chaotic zone can remain stable for a Hubble time.

\begin{figure}
    \centering
    \includegraphics[width=\columnwidth]{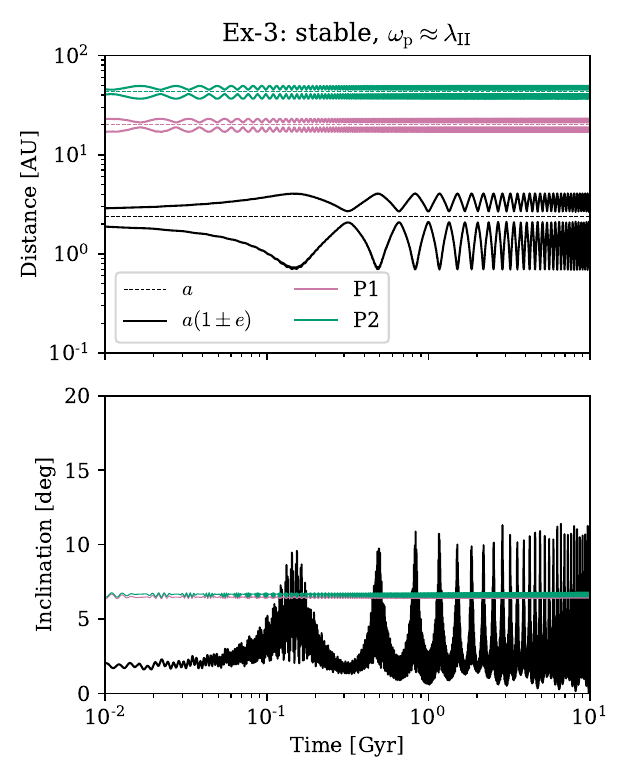}
    \caption{Same as Figs.\ \ref{fig:history_Ex1} and \ref{fig:history_Ex2}, but for run Ex-3. 
    The particle's initial state is $a_{\rm p} = 2.3812 \AU$, $e_{\rm p} = 0.1749$, $I_{\rm p} = 1\fdg8566$, $\varpi_{\rm p} = 180\fdg4930$, $\Omega_{\rm p} = 179\fdg6732$. 
    The integration lasted the full $10.0 \Gyr$; the particle reaches a maximum eccentricity of $0.7111$.}
    \label{fig:history_Ex3}
\end{figure}

Finally, we include run Ex-3 (Fig.\ \ref{fig:history_Ex3}) to emphasize the point that it is the overlap of secular resonances that drives chaotic evolution, rather than resonance itself. 
Here we show the trajectory of a particle within the secular resonance $\omega_{\rm p} \approx \lambda_{\rm II}$, which is well separated from the other major resonances in terms of frequency (Fig.\ \ref{fig:secular_freqs}). 
While the particle's eccentricity indeed becomes large enough that the dynamics are non-linear, the trajectory remains quasiperiodic for $10 \Gyr$.

\subsection{Main experiments} \label{s:expts:main}

\begin{table*}
    \centering
    \begin{tabular}{cccccccccc}
        \hline
        Experiment  & $(m_{1}, m_{2})$ & $e_{1,0}$ & $a_{\rm p}/a_{1}$ & $f_{\rm dis}$ & $\min(t_{\rm dis})$ & $t_{\rm dis,10}$ & $t_{\rm dis,50}$ & $t_{\rm dis,90}$ & $\max(t_{\rm dis})$ \\
        & [$\ME$] & & & & [Gyr] & [Gyr] & [Gyr] & [Gyr] & [Gyr] \\ \hline
        N0 & $30$, $20$ & $0.15$ & $\mathcal{U}(0.237,0.263)$ & $0.810$ & $0.077$ & $0.453$ & $1.213$ & $3.404$ & $9.519$ \\ \hline
        A1 & $95$, $63$ & -- & -- & $0.808$ & $0.028$ & $0.158$ & $0.426$ & $1.433$ & $9.838$ \\
        A2 & $300$, $200$ & -- & -- & $0.853$ & $0.014$ & $0.047$ & $0.134$ & $0.464$ & $9.999$ \\ \hline
        B1 & -- & $0.05$ & -- & $0.096$ & $1.252$ & $2.927$ & $6.146$ & $9.024$ & $9.869$ \\
        B2 & -- & $0.10$ & -- & $0.505$ & $0.202$ & $0.976$ & $2.580$ & $6.338$ & $9.729$ \\
        B3 & -- & $0.25$ & -- & $0.901$ & $0.033$ & $0.137$ & $0.381$ & $1.055$ & $9.768$ \\ \hline
        C1 & -- & -- & $\mathcal{U}(0.297,0.328)$ & $0.996$ & $0.039$ & $0.191$ & $0.693$ & $2.016$ & $8.870$ \\
        C2 & -- & -- & $\mathcal{U}(0.356,0.394)$ & $0.562$ & $0.074$ & $0.341$ & $1.242$ & $4.795$ & $9.646$ \\
        Cw & -- & -- & $\mathcal{U}(0.100,0.450)$ & $0.544$ & $0.065$ & $0.272$ & $0.963$ & $3.667$ & $9.657$ \\
        Cw-cold* & -- & -- & $\mathcal{U}(0.100,0.450)$ & $0.382$ & $0.074$ & $0.295$ & $0.971$ & $2.806$ & $9.956$ \\
        \hline
    \end{tabular}
    \caption{Summary of numerical experiments. 
    A (--) indicates that the entry is identical to the fiducial experiment N0. 
    A tidal disruption threshold $e_{\rm dis} = 0.999$ was assumed in all experiments. 
    The quantity $f_{\rm dis}$ is the fraction of simulated particles that are tidally disrupted within $10 \Gyr$; $t_{\rm dis}$ is the lifetime of a given tidally disrupted particle, while $t_{{\rm dis},p}$ is the $p$-th percentile of the lifetimes of all disrupted particles in a given experiment. 
    (*) Experiment Cw-cold differs from Cw in its initial distributions of $e_{\rm p}$ and $I_{\rm p}$: in Cw-cold, $e_{\rm p} \sim \mathcal{R}(0.03)$ with $\max(e_{\rm p}) = 0.1$ and $I_{\rm p} \sim \mathcal{R}(12^{\circ})$ with $\max(I_{\rm p}) = 10^{\circ}$.}
    \label{tab:tpchaos_ICs}
\end{table*}

We now describe our main series of numerical experiments. 
In each experiment, we conducted 1000 integrations of test particles with semi-major axes drawn from a uniform distribution between $a_{\rm p}/a_{1} = 0.237$ and $0.263$. 
The particle's initial eccentricities, and inclinations, and apsidal and nodal angles were drawn from the distributions listed in the lower part of Table \ref{tab:N0_params} unless otherwise specified. 
We begin with experiment N0, using the outer-planet architecture described in the upper part of Table \ref{tab:N0_params}. 
This serves as the benchmark for our other experiments. 
In most cases, our experiments will differ from N0 in the value of one initial parameter, as described in Table \ref{tab:tpchaos_ICs}. Table \ref{tab:tpchaos_ICs} also summarizes some quantitative results of each experiment, specifically the total fraction of particles that are tidally disrupted in a given experiment ($f_{\rm dis}$) and the typical lifetime ($t_{\rm dis}$) of a tidally disrupted particle.

\begin{figure}
    \centering
    \includegraphics[width=\columnwidth]{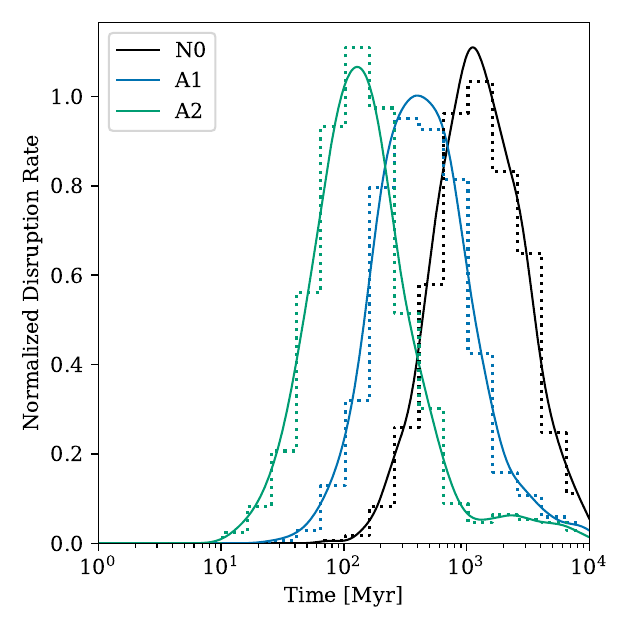}
    \caption{Normalized tidal disruption rate of planetesimals over time in experiments N0, A1, and A2. 
    The outer planets' masses are varied in each experiment, but with a fixed mass ratio $m_{1}/m_{2} = 3/2$: $m_{1} = 30 \ME$ (N0), $30 \times \sqrt{10} \ME \approx 95 \ME$ (A1), and $300 \ME$ (A2). 
    The planetesimal's location is uniformly distributed in the range $a_{\rm p} / a_{1} = [0.237,0.263]$. 
    All other properties of the system are as in Table \ref{tab:N0_params}. 
    Normalized histograms of the times at which tidal disruption events occur are shown as dotted lines, while kernel-density estimates (KDEs) produced from the same data are shown as solid curves.}
    \label{fig:logt_m1}
\end{figure}

Configuration N0 is effective in generating secular chaos. 
Setting the threshold eccentricity for tidal disruption at $e_{\rm dis} = 0.999$, we find that a fraction $f_{\rm dis} = 0.810$ of the simulated test particles were disrupted in $10.0 \Gyr$ or less. 
The 190 `surviving' particles are excluded from the analysis below, since $t_{\rm dis}$ is undefined in these systems. 
As expected, disruption events occur over orders of magnitude in time: To illustrate the distribution, we show the histogram of $t_{\rm dis}$ (with logarithmic bins and normalized to have unit area) as a dotted black contour in Fig.\ \ref{fig:logt_m1}. 
This represents a discretized estimate of the rate $\Gamma_{\rm dis}(t)$ of tidal disruption events in the N0 system as a function of time (or WD cooling age) in normalized units. 
To obtain a continuous version of the same, we perform a Gaussian kernel-density estimation\footnote{As implemented in the {\sc scipy} library \citep{Virtanen+2020}, using Silverman's rule of thumb to select the smoothing bandwidth.} (KDE) on the sequence of values $\{ \log(t_{{\rm dis},i}) \}$. 
The result is shown as a solid black curve in Fig.\ \ref{fig:logt_m1}. 
The distribution has a broad pulse-like profile. The rate of disruption events is essentially zero prior to $t = 100 \Myr$, after which it rises steadily to a maximum around $1.2 \Gyr$ (close to the median value of $t_{\rm dis}$) and then declines. 
The disruption rate is small but finite at $t = 10.0 \Gyr$, when the integration is halted.

In experiments A1 and A2, we scaled up the outer planets' masses by a factor of $\sqrt{10} \approx 3.16$ and $10$, respectively, keeping the mass ratio $m_{2}/m_{1}$ fixed. 
The resulting distributions of $t_{\rm dis}$ are also shown in Fig.\ \ref{fig:logt_m1}, both as histograms and as KDEs. 
(In subsequent figures, we will show only the KDE for visual clarity.) 
One can clearly see that the effect of uniformly scaling the planet masses is to move the center of the distribution while preserving its shape (when represented on a logarithmic axis). 
The median values of the A1 and A2 distributions are roughly $432 \Myr$ and $135 \Myr$, respectively. 
When we compare with the median from N0, we find good agreement with the scaling $t_{\rm dis} \propto m_{1,2}^{-1}$ predicted by Eq.\ (\ref{eq:tdiff}). 
The fraction of planetesimals that are disrupted is similar to N0: $f_{\rm dis} = 0.808$ for A1 and $f_{\rm dis} = 0.853$ for A2.

From the results of A2, one can see that the tidal disruption rate has an extended tail at late times, although the great majority of disruptions occur in the main `pulse' before $t = 1.0 \Gyr$. 
This tail presumably would emerge in the results of N0 and A1 if we were to extend our integrations by a few orders of magnitude in time. 
At late times, the disruption rate declines roughly as $1/t$, which looks like a flat distribution in the logarithmic representation of Fig.\ \ref{fig:logt_m1}. 
This mirrors the results of \citet{PM2017}, who examined the tidal disruption rate over time produced by the Lidov--Kozai mechanism in a stellar binary system. 
We find that the rate of disruption events is well approximated by a log-normal distribution,
\begin{equation} \label{eq:TDE_rate_lognormal_1}
    \Gamma_{\rm dis}(t) \propto \frac{1}{t} \exp\left[ - \frac{\left( \ln{t} - \mu \right)^{2}}{2 \sigma^{2}} \right],
\end{equation}
for $t \lesssim t_{\mu} \equiv \exp\mu$; and by $\Gamma(t) \propto 1/t$ for $t \gg t_{\mu}$. 
The quantity $t_{\mu}$ corresponds closely to the median value of $t_{\rm dis}$ in our numerical experiments, although it can be somewhat smaller for $t_{\mu} \ll 10.0 \Gyr$, since the tail of the distribution is important in such cases. 

\begin{figure}
    \centering
    \includegraphics[width=\columnwidth]{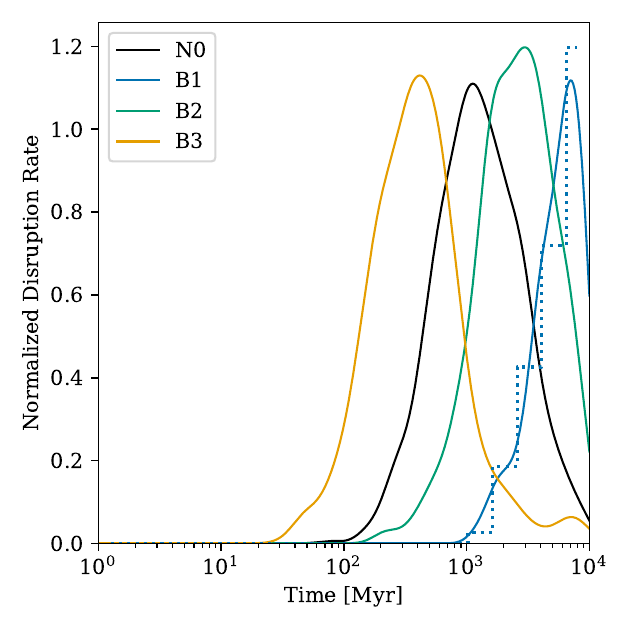}
    \caption{Same as Fig.\ \ref{fig:logt_m1} for experiments B1, B2, and B3. 
    Only the KDE curves are shown for clarity, except for B1. 
    The inner planet's initial eccentricity is varied: $e_{1} = 0.15$ (N0), $0.05$ (B1), $0.1$ (B2), and $0.25$ (B3). 
    All other properties of the system are as in Table \ref{tab:N0_params}.}
    \label{fig:logt_e1}
\end{figure}

In experiments B1, B2, and B3, we varied the initial eccentricity $e_{1}$ of the inner planet, setting it to $0.05$, $0.10$, and $0.25$ respectively; the initial eccentricity of the outer perturber was fixed at $e_{2} = 0.05$ in all cases, as was the perturbers' mutual inclination of $13^{\circ}$. 
Fig.\ \ref{fig:logt_e1} shows the results of these experiments. 
The distribution of disruption times is broadly similar to the previous three experiments. 
We see that the distribution shifts significantly towards shorter $t_{\rm dis}$ with increasing $e_{1}$. 
Quantitatively, the shift is roughly consistent with the power law $t_{\rm diff} \propto e_{1}^{-1}$ predicted by Eq.\ (\ref{eq:tdiff}). 
Finally, we note that in Experiment B1, the disruption rate appears to decline abruptly at late times. 
This is a smoothing artefact arising from the fact that we end our simulations at $t = 10.0 \Gyr$; a histogram shows that the rate of disruption events is actually still increasing at late times in this experiment.

Experiments B1 and B2 stand out in Table \ref{tab:tpchaos_ICs} for having significantly smaller fractions of disrupted particles than N0 -- $f_{\rm dis} = 0.096$ and $0.505$, respectively. 
This lower efficiency of secular chaos reflects two factors. 
First, the chaotic diffusion rate is lower in B1 and B2, so fewer particles reach the tidal disruption threshold within 10 Gyr; if we were to extend the integrations beyond 10 Gyr, we might find higher values of $f_{\rm dis}$. 
Second, the chaotic zone in phase space is expected to expand or contract in proportion to the secular mode amplitude $|c_{\rm I}| \sim e_{1,0}$ \citep{LW2011}. 
The same trends cause B3, with more dynamically active perturbers than N0 and correspondingly faster chaotic diffusion, to have a larger $f_{\rm dis} = 0.901$.

\begin{figure}
    \centering
    \includegraphics[width=\columnwidth]{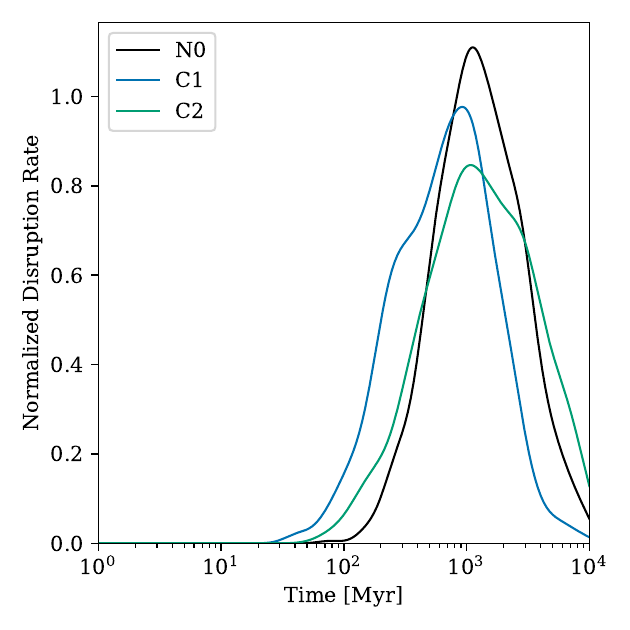}
    \caption{Same as Fig.\ \ref{fig:logt_m1} for experiments C1 and C2. 
    The semi-major axis ratio $a_{\rm p}/a_{1}$ is varied: the values of $a_{\rm p}/a_{1}$ are sampled uniformly in the ranges $[0.237,0.263]$ (N0), $[0.297,0.323]$ (C1), and $[0.356,0.394]$ (C2). 
    All other properties of the system are as in Table \ref{tab:N0_params}.}
    \label{fig:logt_alpha1}
\end{figure}

In experiments C1 and C2, we varied the ratio $a_{\rm p}/a_{1}$, holding fixed the planets' semi-major axes $a_{1}$ and $a_{2}$. 
Fig.\ \ref{fig:logt_alpha1} shows the results of these experiments. 
In this case, the numerical results deviate significantly from predicted scaling law of Eq.\ (\ref{eq:tdiff}). 
While the linear secular theory suggests $t_{\rm diff} \propto (a_{\rm p} / a_{1})^{-4}$, the numerical results show that the dependence on $(a_{\rm p}/a_{1})$ is non-monotonic. 
Increasing from $a_{\rm p}/a_{1} = 0.25$ in experiment N0 to $a_{\rm p}/a_{1} = 0.313$ in C1 decreased the median of $t_{\rm dis}$ a factor of $\sim 2$. 
However, increasing further to $a_{\rm p}/a_{1} = 0.375$ (C2) essentially returned the median to its former value. 
On the other hand, changing $a_{\rm p}/a_{1}$ appears to change shape of the distribution somewhat (unlike variations of the planets' masses or eccentricities). 

The non-monotonic dependence of the tidal disruption rate on $a_{\rm p}/a_{1}$ is astrophysically interesting because a real planetesimal belt may have significant radial extent. 
To probe the radial structure of the chaotic zone, we carried out experiment Cw (`w' for `wide'), in which we conduct $1200$ runs with $a_{\rm p}/a_{1}$ drawn from a uniform distribution between $0.1$ and $0.45$; all other initial conditions were generated as in N0. 
In the upper panel of Fig.\ \ref{fig:lifetimes_Cw}, we show the lifetimes of individual particles in this experiment, colour-coded by whether they were tidally disrupted within $10 \Gyr$. 
The lifetimes of the disrupted particles trace the non-monotonic structure hinted at by experiments C1 and C2, albeit with coarse resolution. 
In the lower panel, we show the fraction of particles that are disrupted within bins of width $0.025$. 
One can clearly see that the chaotic zone is radially localized in the vicinity of the linear secular resonances $\gamma=\lambda_{\rm I}$ and $\gamma=|\lambda_{0}|$ (indicated by the vertical lines near $a_{\rm p}/a_{1} = 0.3$). 
Between $a_{\rm p}/a_{1} = 0.25$ and $0.35$, over 90 per cent of test particles are tidally disrupted. 
The typical particle lifetime is also at a minimum in this interval. 
However, the chaotic zone also has significant `wings' at smaller and large $a_{\rm p}/a_{1}$: the disrupted fraction goes above $0.1$ near $a_{\rm p}/a_{1} = 0.15$ and above $0.3$ near $a_{\rm p}/a_{1} = 0.45$. 
The total fraction of particles disrupted (for a uniform distribution of the semi-major axis) is $0.554$.

A major reason for the large radial width of the chaotic zone in experiment Cw is that we have assumed a dynamically hot distribution of initial eccentricities and inclinations for the test particles (Tab.\ \ref{tab:tpchaos_ICs}). 
As Fig.\ \ref{fig:secular_freqs} shows, a particle with a high eccentricity (or inclination from the invariable plane) can experience non-linear secular resonances over a wide range of $a_{\rm p}$. 
One might therefore ask whether a dynamically colder ensemble of test particles would be less susceptible to chaotic evolution. 
To test this, we conduct a final numerical experiment, called `Cw-cold'. 
The setup is identical to Cw expect that the r.m.s.\ initial eccentricities and inclinations of the test particles are reduced by a factor of 4 (i.e.\ to $0.03$ and $3^{\circ}$, respectively) and the Rayleigh distributions are truncated at eccentricity $0.1$ and inclination $10^{\circ}$. 
The results are displayed in Fig.\ \ref{fig:lifetimes_Cw-cold}. 
As expected, we see that the chaotic zone is significantly narrower in terms of semi-major axis and that fewer particles are tidally disrupted within $10 \Gyr$. 
However, the Cw-cold chaotic zone is still fairly extensive (roughly between $a_{\rm p}/a_{1} = 0.25$ and $0.4$) and about 38 per cent of all particles are disrupted. 
The typical particle lifetime within the chaotic zone is essentially unchanged. 
This suggests that the eccentricity and inclination dispersions of a planetesimal belt play a less important role in determining its susceptibility to secular chaos than its proximity to the system's linear secular resonances.

The range of $a_{\rm p}/a_{1}$ we have tested roughly encompasses the radial extent of the chaotic zone due to secular effects. 
For $a_{\rm p}/a_{1} \gtrsim 0.45$, low-order mean-motion resonances occur that can lead to additional effects not captured by the ring-averaging method \citep[e.g.,][]{PML2017}. 
Moreover, for $a_{\rm p}/a_{1} \gtrsim 0.5$ the excitation to $e_{\rm p} \simeq 1$ could lead to close encounters with the inner planet, meaning that the orbital evolution ceases to be secular. 
On the other hand, for $a_{\rm p}/a_{1} \lesssim 0.1$ we exit the region of overlapping secular resonances and no longer see chaotic secular evolution (Fig.\ \ref{fig:history_Ex3}). 
The chaotic range of $a_{\rm p}/a_{1}$ could be changed by altering the perturbers' mass ratio $m_{2}/m_{1}$ or spacing $a_{1}/a_{2}$, thereby moving the linear secular resonances; however, this probably would not change the qualitative behaviours we see in Figs.\ \ref{fig:lifetimes_Cw} and \ref{fig:lifetimes_Cw-cold}.

\section{Discussion} \label{s:discuss}

\subsection{WD pollution from secular chaos} \label{s:discuss:WDpol}

\subsubsection{Metal accretion rate vs.\ cooling age} \label{s:discuss:Mdot_vs_age}

As discussed in Section \ref{s:intro}, dynamical models of WD pollution are constrained mainly by the observed distribution of the metal accretion rate ($\dot{M}_{Z}$) as a function of cooling age. 
Our numerical experiments have provided an analytic approximation of the tidal disruption rate of planetesimals over time; for a given model of the planetesimal belt and the outer-planet architecture, we can compute the expected rate at which mass is delivered to the WD's Roche limit as a function of time. 
Observations suggest that the debris of a disrupted planetesimal quickly forms an accretion disc near the Roche limit; the metal accretion rate onto the WD is determined by the balance of disc physics and the supply of debris.

First, let us consider a simplified model of the mass budget in a polluted WD's accretion disc. 
We assume that, when a planetesimal is tidally disrupted, its mass is immediately incorporated into a disc. 
We also assume that the WD accretes matter from the disc on a fixed time-scale $\tau_{\rm acc}$. 
Let the rate at which mass is supplied externally to that disc be $\dot{M}_{\rm ext}(t)$; this is directly related to the tidal disruption rate. 
The disc's total mass $M_{\rm d}(t)$ obeys the equation
\begin{equation}
    \frac{\dif M_{\rm d}}{\dif t} = - \frac{M_{\rm d}}{\tau_{\rm acc}} + \dot{M}_{\rm ext}(t).
\end{equation}
If $M_{\rm d}(0) = 0$, the solution is
\begin{equation}
    M_{\rm d}(t) = e^{-t/\tau_{\rm acc}} \int_{0}^{t} \dot{M}_{\rm ext}(t') e^{t'/\tau_{\rm acc}} \, \dif t'.
\end{equation}
An important special case is when $\dot{M}_{\rm ext}(t)$ varies on time-scales much longer than $\tau_{\rm acc}$. 
The system relaxes to a steady state for $t \gg \tau_{\rm acc}$, and the WD's accretion rate is given by
\begin{equation}
    \dot{M}_{Z}(t) \simeq \frac{M_{\rm d}(t)}{\tau_{\rm acc}} \simeq \dot{M}_{\rm ext}(t).
\end{equation}
Thus, the observed accretion rate of a given WD reflects the average supply rate of planetesimals, which can be directly related to the dynamical evolution of external ``planetesimals + planets'' system.

In reality, mass is supplied to the accretion disc stochastically and in discrete amounts. 
The approximation of continuous delivery fails if the average interval $\Delta t_{\rm dis}$ between tidal disruption events is much longer than $\tau_{\rm acc}$. 
In that case, the instantaneous accretion rate can differ from the average supply rate by orders of magnitude and may be either larger or smaller \citep{Wyatt+2014}. 
Studies of WDs with infrared excess emission from compact, dusty discs estimate that the typical disc lifetime is $\sim 10^{4}$--$10^{6} \yr$ (e.g., \citealt{FJZ2009}, \citealt{Girven+2012}; but note that \citealt{Wyatt+2014} argue that it could be much shorter). 
The value of $\Delta t_{\rm dis}$, meanwhile, depends on the external planetesimal reservoir's intrinsic properties and dynamical evolution.

Using our numerical simulations (Section \ref{s:expts}), we have estimated the tidal disruption rate per particle $\Gamma_{\rm dis}(t)$, which gives the fraction $\dif f_{\rm dis}$ of a planetesimal belt initially containing $N_{0} \gg 1$ bodies that is disrupted between $t$ and $t + \dif t$. 
In our numerical experiments, the log-normal component of $\Gamma_{\rm dis}$ accounts for the great majority of disruption events from secular chaos. 
Ignoring the power-law tail at late times, then, we can approximate the TDE rate as (see Eq.\ \ref{eq:TDE_rate_lognormal_1})
\begin{equation} \label{eq:TDE_rate_lognorm}
    \Gamma_{\rm dis}(t;\mu,\sigma) = \frac{f_{\rm dis}}{(2 \pi)^{1/2} \sigma t} \exp\left[ - \frac{\left( \ln{t} - \mu \right)^{2}}{2 \sigma^{2}} \right],
\end{equation}
where $\mu$ and $\sigma$ depend on the planetary system architecture as described in Section \ref{s:expts}. 
The total disrupted fraction $f_{\rm dis}$ depends primarily on the eccentricity and inclination dispersion of the initial belt and perturbers -- in other words, on the fraction of the initial orbits that actually lie in the chaotic zone. 
Referring to experiments N0, A1, and A2 for purposes of illustration, we find $f_{\rm dis} \approx 0.8$--$0.9$, $\sigma \approx 0.4$, and $t_{\mu} \approx 1.0 \Gyr \, (m_{1} / 30 \ME)^{-1}$ (with a fixed $m_{2}/m_{1}$). 
The maximal disruption rate occurs at $t = t_{\mu}$, and thus the total disruption rate around that time is
\begin{equation}
    \mathcal{R}_{\rm max} = N_{0} \Gamma_{\rm dis}(t_{\mu}) = \frac{N_{0} f_{\rm dis}}{(2 \pi)^{1/2} \sigma t_{\mu}}.
\end{equation}
The logarithmic width $\sigma$ varies only by a factor of $\sim 2$ across our various experiments; therefore, the average interval $\Delta t_{\rm dis}$ between disruption events is determined primarily by the combination $t_{\mu} / (N_{0} f_{\rm dis})$:
\begin{align}
    \Delta t_{\rm dis} &= \frac{1}{\mathcal{R}_{\rm max}} \nonumber \\
    &\approx \frac{10^{6} \yr}{f_{\rm dis}} \left( \frac{t_{\mu}}{1.0 \Gyr} \right) \left( \frac{\sigma}{0.4} \right) \left( \frac{N_{0}}{10^{3}} \right)^{-1}. \label{eq:Delta_tdis}
\end{align}
The reference value $N_{0} = 10^{3}$ corresponds to the number of test-particle integrations carried out in each of our experiments. 
Fortuitously, it also represents an approximate upper limit on the number of asteroids in the present-day Solar System main belt that could plausibly survive the late stages of stellar evolution. 
We elaborate on this point in Section \ref{s:disc:preWD:late}.

Note that Eq.\ (\ref{eq:Delta_tdis}) gives the {\it minimum} of $\Delta t_{\rm dis}$ over the $\sim 10 \Gyr$ lifetime of the system; for cooling ages much less or much greater than $t_{\mu}$, $\Delta t_{\rm dis}$ would be longer. 
Considering the range of typical disc lifetimes to be $\sim 10^{4}$--$10^{6} \yr$, we conclude that the supply of mass to the accretion disk mostly cannot be approximated as continuous unless $t_{\mu} / (N_{0} f_{\rm dis}) \ll 1 \Myr$. 
This conclusion is strengthened if disc lifetimes are actually of the order of decades or centuries, as suggested by \citet{Wyatt+2014}; or if the number of surviving planetesimals is much less than the reference value in Eq.\ (\ref{eq:Delta_tdis}) (see Section \ref{s:disc:preWD:late}). 
Because of these considerations, we must be cautious in trying to relate the tidal disruption rate from our simulations to the measured accretion rates of polluted WDs.

Various studies have identified trends of accretion rates versus cooling age in large samples of polluted WDs. 
We briefly summarize these results in the next paragraph before comparing them with our numerical experiments. 
We divide the population of polluted WDs into three cohorts by cooling age: `young' WDs with cooling ages $< 0.1 \Gyr$; `middle-aged' WDs between $0.1$ and $1 \Gyr$; and `old' WDs with cooling ages $> 1.0 \Gyr$.

\begin{figure}
    \centering
    \includegraphics[width=\textwidth]{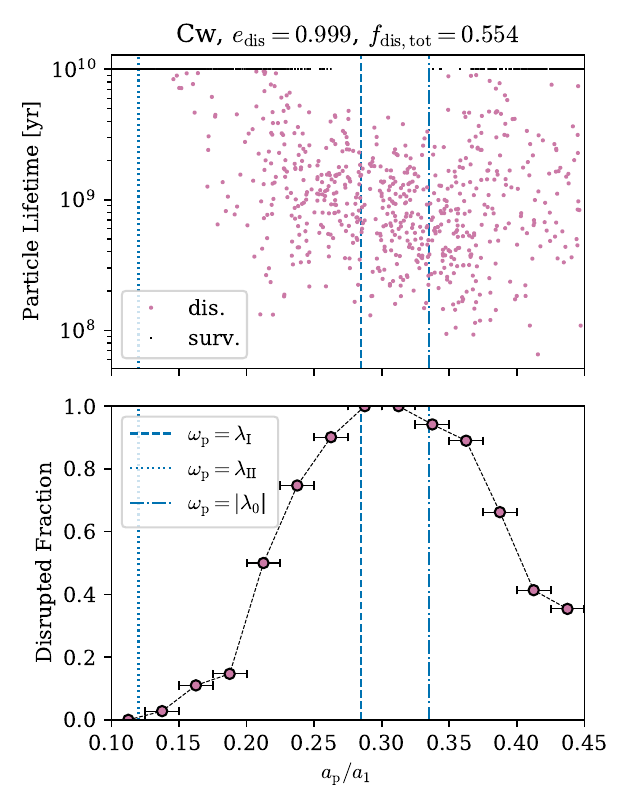}
    \caption{Results of numerical experiment Cw (see Table \ref{tab:tpchaos_ICs}). 
    In both panels, vertical lines show the locations of linear secular resonances. 
    {\it Upper:} Scatter plot of test-particle lifetimes. 
    Particles that reached the disruption threshold $e_{\rm dis} = 0.999$ within $10 \Gyr$ are shown as purple dots. 
    Particles that were not disrupted within $10 \Gyr$ are shown as black pixels along the horizontal line with lifetime $= 10 \Gyr$. 
    {\it Lower:} The fraction of test particles disrupted within $10 \Gyr$, binned in increments of $\Delta(a_{\rm p}/a_{1}) = 0.025$ (indicated by the horizontal error bars).}
    \label{fig:lifetimes_Cw}
\end{figure}

\begin{figure}
    \centering
    \includegraphics[width=\textwidth]{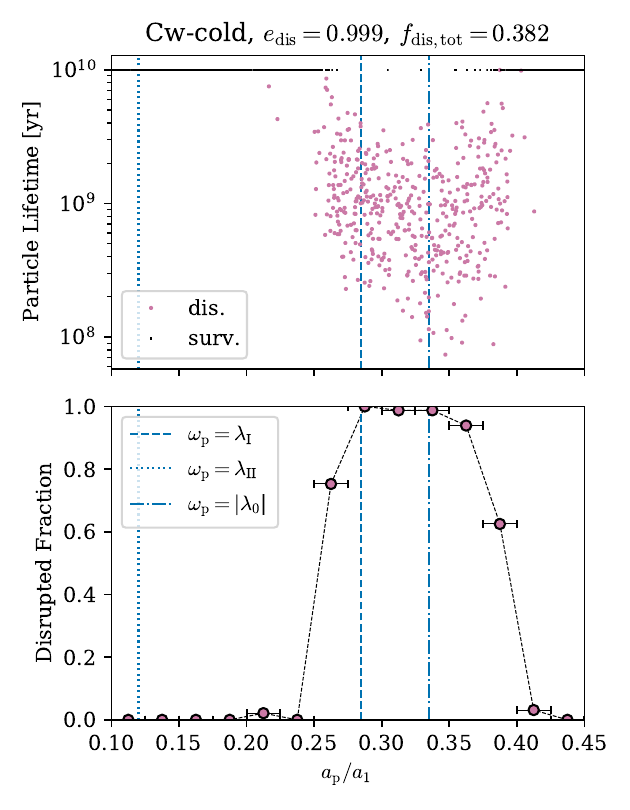}
    \caption{Same as Fig.\ \ref{fig:lifetimes_Cw} for experiment Cw-cold.}
    \label{fig:lifetimes_Cw-cold}
\end{figure}

\begin{figure*}
    \centering
    \includegraphics[width=0.55\textwidth]{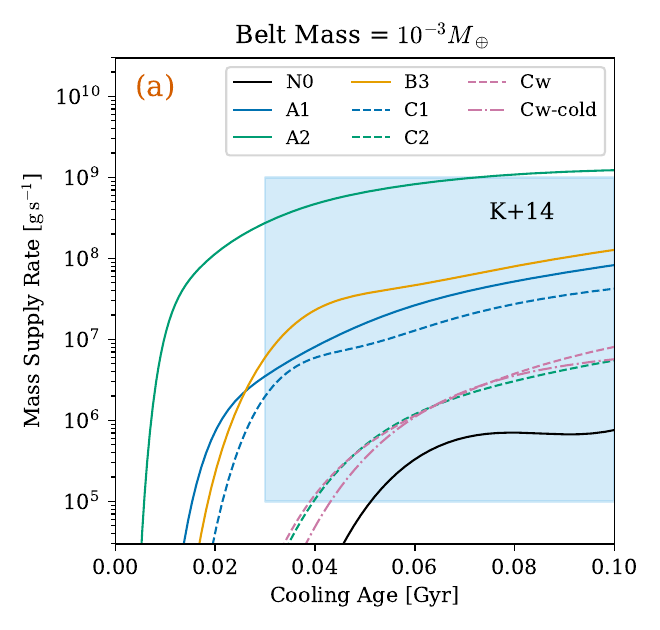}
    \includegraphics[width=\textwidth]{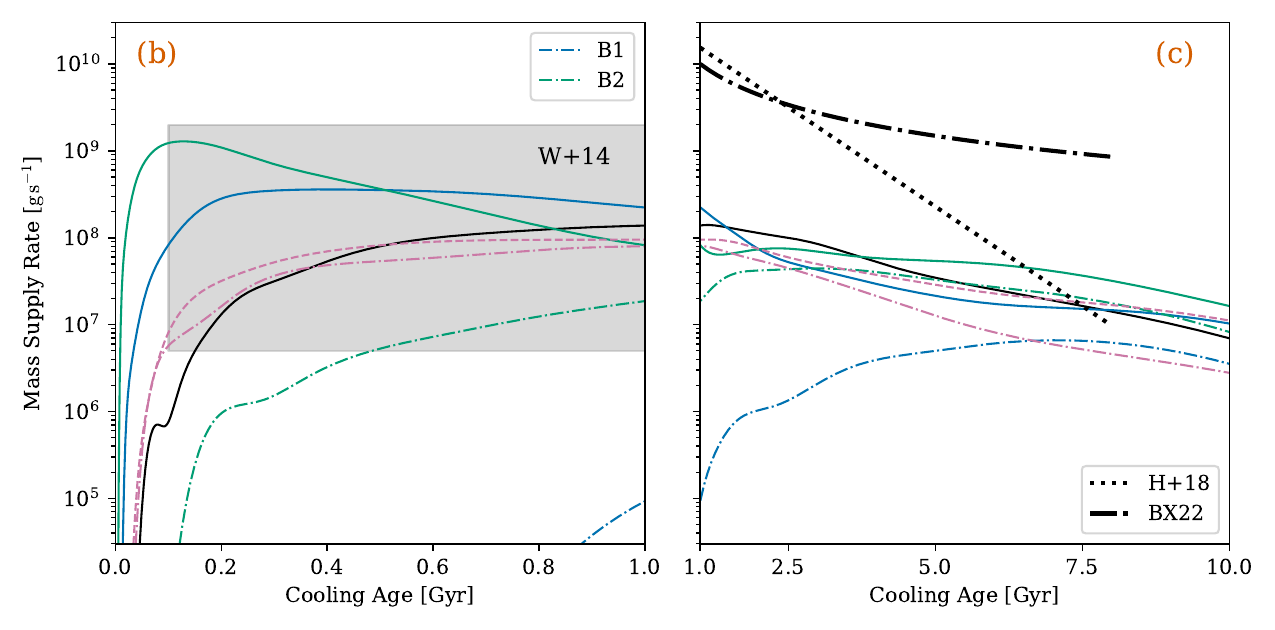}
    \caption{Metal accretion rates versus cooling age for polluted WDs. 
    Different curves show the estimated mass supply rate $\dot{M}_{\rm ext}$ as a function of time for several of our numerical experiments. 
    An initial belt mass of $M_{\rm b0} = 10^{-3} \ME$ is assumed throughout. 
    {\it (a)} Accretion rates for young ($< 0.1 \Gyr$) WDs. 
    The light-blue shaded region indicates the approximate range of metal accretion rates $\dot{M}_{Z}$ calculated by \citet{KGF2014} for a sample of WDs with ages between $30$ and $100 \Myr$, assuming bulk Earth abundances. 
    {\it (b)} Same as (a) for middle-aged ($0.1$--$1 \Gyr$) WDs. 
    Several curves are continuations of those displayed in (a), but for visual clarity only N0, A1, and A2 are included. 
    We also show cases B1 and B2, which were not visible in (a), as dot-dashed coloured curves. 
    The grey shaded region indicates the range of supply rates for WDs with cooling ages between $0.1$ and $1 \Gyr$ modelled by \citep{Wyatt+2014}. 
    The accretion rates of the individual polluted WDs studied by \citet{KGF2014} in this age bracket largely fall within the same range. 
    {\it (c)} Same as (a) and (b) for old ($> 1 \Gyr$) WDs. 
    The dotted and dot-dashed black curves show the {\it upper envelope} of the accretion-rate distribution identified by \citet{HGK2018} and \citet{BX2022}, respectively.}
    \label{fig:Mext_age}
\end{figure*}

For young WDs, \citet{KGF2014} found that external metal accretion from rocky parent bodies apparently begins at cooling ages of $\sim 30$--$50 \Myr$. 
Among young WDs for which radiative levitation is ruled out by \citeauthor{KGF2014}, the inferred metal accretion rates were between $\sim 10^{5}$ and $10^{9} \, {\rm g \, s^{-1}}$. 
For middle-aged WDs, the average supply rate appears to be constant \citep[e.g.,][]{KGF2014,Xu+2019b,BX2022}. 
\citet{Wyatt+2014} found that the observations are consistent with an underlying log-normal distribution of the averaged supply rate $\dot{M}_{\rm ext}$ with median $10^{8} \, {\rm g \, s^{-1}}$ and standard deviation $1.3 \, {\rm dex}$. 
This is broadly consistent with accretion rates for younger WDs \citep{KGF2014}. 
Finally, for old WDs, there is not yet consensus on the observed trend of age versus accretion rate: \citet{HGK2018} find that the largest accretion rate observed at a given cooling age declines exponentially with an e-folding time of $0.95 \Gyr$. 
On the other hand, \citet{BX2022} find that the upper envelope declines by no more than $1 \, {\rm dex}$ for cooling ages between $1$ and $8 \Gyr$. Finally, the lower limit of the observed accretion rates is consistently $\sim 10^{5} \, {\rm g \, s^{-1}}$ for most WDs, since it is determined by the observational detection threshold \citep[e.g.,][]{KGF2014,BX2022}. 
Despite the uncertainties, these results provide useful benchmarks to evaluate whether our dynamical scenario can broadly reproduce observations.

In Figure \ref{fig:Mext_age}a, we compare the metal supply rate predicted by several of our numerical experiments to the range of metal accretion rates among young WDs, assuming an initial planetesimal belt mass $M_{\rm b0} = 10^{-3} \ME$ (roughly twice the present mass of the Solar System main belt). 
In general, we find a precipitous rise in our estimated metal supply rate, $\dot{M}_{\rm ext} = M_{\rm b0} \Gamma_{\rm dis}(t)$, followed by a levelling off or slower increase. 
Several experiments reproduce the observed onset of accretion after $\sim 20 \Myr$ fairly well, and $\dot{M}_{\rm ext}(t)$ is mostly within the observed range of accretion rates for young WDs \citep{KGF2014}. 
We emphasize that $\dot{M}_{\rm ext}$ represents the pollution rate expected for continual delivery of small, equal-mass planetesimals, which is not expected physically (see above) but remains useful for illustrative purposes. 
The actual pollution rate observed at a given WD at a given cooling age can vary by a few orders of magnitude from $\dot{M}_{\rm ext}$.

Experiment A2 notably predicts an early onset of accretion ($\sim 10 \Myr$) compared to the observations and generally errs on the side of too large an accretion rate for young WDs. 
The latter issue can be ameliorated somewhat by reducing $M_{\rm b0}$ by an order of magnitude (i.e.\ to $\sim 10^{-4} \ME$). 
However, the former cannot be solved in this way because it is reflects the rate of chaotic diffusion in this system, which in turn reflects the configuration of outer planets. 
On the other hand, the rapid dynamical evolution of A2 may be advantageous for explaining other observations of the planetary systems of young WDs (see Section \ref{s:discuss:HEM:planets}). 
We also have not accounted for the physics of accreting planetary debris around very young, hot, luminous WDs. 
For example, sublimation of dust grains at or beyond the Roche limit around a hot WD may delay or prevent accretion of tidally disrupted debris until after the WD has cooled sufficiently \citep{Steckloff+2021}. 
This may allow systems like A2 to remain consistent with the observations despite their apparently large accretion rates at early times.

We now compare our simulation results with the accretion rates of middle-aged and old polluted WDs in Figures \ref{fig:Mext_age}b and c, respectively. 
For visual clarity, we omit the curves for experiments B3, C1, and C2 (note that B3 and C1 are quantitatively similar to A1, while C2 is similar to N0). 
For middle-aged WDs, we see that curves N0, A1, and A2 mostly agree with the observations (the shaded region, based mainly on \citealt{Wyatt+2014} but consistent with \citealt{KGF2014} and \citealt{BX2022}). 
Experiment B2 tends to under-predict the accretion rates of middle-aged WDs somewhat, while B1 falls far short. 
Increasing $M_{\rm b0}$ by an order of magnitude (i.e.\ to $10^{-2} \ME$) can bring B2 into better agreement. 
However, B1 cannot be reconciled with the middle-aged WD population because the required planetesimal belt mass is in excess of $1 \ME$; this is large enough for the self-gravity of the belt to suppress secular chaos (see Section \ref{s:disc:preWD:selfgravity}).

For old WDs, our estimated $\dot{M}_{\rm ext}$ are fully compatible with the tentative upper envelope of accretion rates determined by \citet{BX2022}. 
Interestingly, our results naturally reproduce the gradual downward trend identified in that study: roughly $1 \, {\rm dex}$ from $1$ to $8 \Gyr$. 
Since the trend line of \citet{BX2022} lies well above our simulation results, a larger $M_{\rm b0}$ than our fiducial value ($10^{-3} \ME$) may be needed to reproduce the largest accretion rates reported for old WDs ($\sim 10^{9}$--$10^{10} \, {\rm g \, s^{-1}}$). 
However, due to stochastic accretion effects, the instantaneous accretion rate for a given WD can vary by a few orders of magnitude about $\dot{M}_{\rm ext}$ \citep{Wyatt+2014}. 
On the other hand, if we adopt the upper envelope of \citet{HGK2018} as our observational benchmark, then all of our simulations predict too gradual a decrease of accretion rates for old WDs. 
The slow decay of $\dot{M}_{\rm ext}$ at late times is an intrinsic feature of our dynamical scenario because it reflects the diffusive nature of secular chaos.

An interesting observation about Figs.\ \ref{fig:Mext_age}bc is that the exact properties of the planetesimal belt -- particularly its radial width and its initial eccentricity and inclination dispersion -- do not appear to greatly affect the accretion rate over time. 
One difference is that, in Cw and Cw-cold, the accretion rate rises slightly faster at early times than in N0 (because many planetesimals occupy the deepest part of the chaotic zone at early times; see Figs.\ \ref{fig:lifetimes_Cw} and \ref{fig:lifetimes_Cw-cold}). 
At late times, the accretion rate is somewhat lower in Cw-cold because the chaotic zone contains fewer planetesimals overall compared to N0 and Cw (i.e.\ $f_{\rm dis}$ is lower).

On the whole, Fig.\ \ref{fig:Mext_age} shows that secular chaos driven by planets larger than $\sim 10 \ME$ beyond $\sim 10 \AU$ can sustain metal accretion rates consistent with observations across most of the WD cooling sequence. 
This depends on the assumed configuration of outer planets and the mass of the planetesimal belt. 
Some configurations perform better than others: For an initial belt mass of $10^{-3} \ME$, B1 predicts no observable metal accretion until after $1 \Gyr$ of cooling, and even then $\dot{M}_{\rm ext}$ remains relatively low. 
On the other hand, A2 with the same belt mass matches the middle-aged and old WD populations but predicts the onset of significant metal accretion after just $\sim 10 \Myr$, which is not observed \citep[but see][]{Steckloff+2021}. 
Three configurations that perform especially well are A1, C1, and B3: all of them are approximately consistent with the observed accretion rates across the full range of cooling ages we have considered. 
We note that our assumed initial planetesimal belt mass of $10^{-3} \ME$ is only twice the mass of the asteroid belt in the present-day Solar System. 
In some cases, increasing or decreasing the belt mass by a factor of $\sim 10$ somewhat improves the agreement with observations. 
The amount of mass that is expected to remain in a planetesimal belt after stellar evolution is uncertain, although some theoretical limits exist for the post-MS survival of planetesimals as a function of size, location, and stellar mass \citep[e.g.,][]{BW2010,VJG2014,Martin+2020}. 
In any case, it is encouraging that the accretion rates produced by secular chaos broadly agree with observations even when the total mass of surviving planetesimals in the chaotic zone is relatively low. 
This reflects the high fraction of test particles that achieve extreme eccentricities through secular chaos.

\subsubsection{Comparison with related works} \label{s:discuss:WDpol:comparison}

The possibility of driving WD pollution through secular resonances with giant planets was previously suggested by \citet{Smallwood+2018,Smallwood+2021}, who studied the pollution induced by the $\nu_{6}$ resonance in the asteroid belt of a hypothetical evolved Solar System. 
We have extended that study in several ways. Our use of the ring-averaging method has allowed us to simulate the secular evolution of a belt--perturber system over $10 \Gyr$, while the $N$-body simulations of \citet{Smallwood+2018,Smallwood+2021} last only $50$ and $100 \Myr$, respectively. 
We have also investigated the effect of varying the masses and orbital elements of the perturbing giant planets, which allows us to extrapolate our results to different system architectures. 
Finally, we find a much higher fraction of tidally disrupted planetesimals in our simulations and a correspondingly larger accretion rate onto the WD. 
This is presumably because we have assumed outer planets that are more eccentric and more inclined than Jupiter and Saturn, similar to the population of long-period extrasolar giant planets.

\citet*{Li+2022} have also studied WD pollution in the evolved Solar System using $N$-body simulations to a cooling age of $2 \Gyr$. 
Their ``count-based'' estimate (which is most directly comparable to our $\dot{M}_{\rm ext}$) indicates that the metal accretion rate of the solar WD due to main-belt asteroids decays roughly as a single power law from $\sim 10^{8} \, {\rm g \, s^{-1}}$ at a cooling age of $1 \Myr$ to $\sim 10^{6} \, {\rm g \, s^{-1}}$ at $2 \Gyr$. 
This differs significantly from the prediction of our secular-chaos scenario in several respects. 
We predict a significantly higher accretion rate across most of the WD cooling sequence even though our initial planetesimal belt is only a factor of 2 more massive. 
\citet{Li+2022} do not find an abrupt onset of pollution at a characteristic cooling age (cf.\ Fig.\ \ref{fig:Mext_age}a) and predict a more rapid decline of the accretion rate at late times (cf.\ Figs.\ \ref{fig:Mext_age}bc). 
These differences could arise from differences between our fiducial planetary system architecture and the Solar System. 
In particular, the main belt is closer to Jupiter ($0.45 \lesssim a_{\rm p}/a_{1} \lesssim 0.8$) than the planetesimal belt is to the inner planet in our scenario ($0.15 \lesssim a_{\rm p}/a_{1} \lesssim 0.45$). 
This means both that close encounters are possible and that low-order mean-motion resonances can affect asteroids in the Solar System. 
Again, the secular resonances located near the main belt have smaller chaotic zones because Jupiter and Saturn have modest eccentricities and mutual inclinations compared to the planets in our simulations.

\citet{Mustill+2018} studied planet--planet scattering in systems of three or more low-mass ($\sim 1$--$30 \ME$) planets orbiting a WD beyond $\sim 10 \AU$. 
They found that scattering has a strong disruptive effect on neighbouring planetesimal belts and that planetesimal pollution can be sustained in these systems because the instabilities can last for many Gyr. 
Some of our predictions are similar to theirs: for instance, both scenarios can reproduce the observed onset of WD pollution at cooling ages $\sim 30$--$50 \Myr$. 
In both cases, this reflects a `ramp-up' time to excite planetesimal eccentricities to large values. 
For us, the ramp-up time is proportional to the chaotic diffusion time-scale. 
For \citet{Mustill+2018}, it reflects the delay between the expansion of planetary orbits during stellar evolution and the onset of planet--planet scattering. 
However, we make markedly different predictions for old WDs: Their simulations predict an exponentially declining trend of accretion rate versus cooling age with an e-folding time of $\sim 1 \Gyr$. 
This is consistent with the upper envelope of the observed accretion rates as fitted by \citet{HGK2018}. 
Our simulations are more consistent with the more gradual trend fitted by \citet{BX2022}, which \citet{Mustill+2018} do not reproduce.

Our results demonstrate that secular chaos is a viable alternative to the low-mass planet--planet scattering scenario of \citet{Mustill+2018} as a way to sustain WD pollution for many Gyr. 
While the cooling age at which the pollution rate is greatest depends on the mass and architecture of the perturbers, we find that pollution rates are sustained over $10 \Gyr$ in all cases. 
Notably, we find that Jupiter-mass planets can contribute to the pollution of WDs much older than $1 \Gyr$, in contrast with previous findings that such planets deplete planetesimal reservoirs too quickly for this occur \citep[e.g.,][]{DWS2012,FH2014,Mustill+2018}. 
The main reason for this is the diffusive nature of secular chaos, which causes tidal disruption events to occur over several orders of magnitude in time. 
To a lesser extent, it may also be because our simulated planetary systems are somewhat larger in terms of semi-major axis than those of some previous studies \citep[e.g.,][]{DWS2012,FH2014} and consequently have a longer dynamical timescale ($\propto a^{3}/m$, Eqs.\ \ref{eq:om_jk}--\ref{eq:nu_jk}).

In Section \ref{s:discuss:Mdot_vs_age}, we pointed out that our scenario requires a relatively low-mass planetesimal belt ($\sim 10^{-3} \ME$ in total) in order to reproduce the observed metal accretion rates of WDs. 
On the other hand, previous studies of WD pollution through direct scattering of planetesimals \citep{FH2014,Mustill+2018} or low-order mean-motion resonances \citep{DWS2012} have found that substantially more massive belts (up to $\sim \ME$) are required to reproduce observations. 
This is because, in those scenarios, only a small fraction of the planetesimal belt is tidally disrupted by the WD, with most bodies being ejected entirely. 
In our scenario, planetesimals that experience chaotic diffusion cannot be ejected from the system because (i) they cannot exchange energy with the planets under the secular approximation; and (ii) their orbits are well inside the planetary orbits, i.e.\ $a_{\rm p} (1 + e_{\rm p}) \lesssim a_{1} (1 - e_{1})$, preventing close encounters. 
Thus, nearly all of the available mass of planetesimals inside the chaotic zone is eventually accreted by the WD.

\subsubsection{Prospects for detecting outer planets} \label{s:discuss:WDpol:detect}

More than 25 per cent of solitary WDs are polluted by planetary material \citep[e.g.,][]{Zuckerman+2010,KGF2014}. 
If secular chaos is a major dynamical channel for pollution, it follows that a significant fraction of the WD population possesses at least two planets with masses $\gtrsim 10 \ME$ on long-term stable orbits beyond $\sim 10 \AU$, with moderate eccentricities and mutual inclinations. 
By extension, a similar fraction of F- and A-type main-sequence stars (the progenitors of polluted WDs) must have similar planetary systems with smaller orbits. 
A reservoir of planetesimals close to the chaotic zone would also be required, which implies that the approximate configuration of the main asteroid belt and outer Solar System planets is commonplace.

Could this hypothetical planet population ever be detected? Direct detection of extrasolar planets beyond $\sim 10 \AU$ is difficult for main-sequence host stars, let alone WDs. 
None the less, \citet{Schreiber+2019} have reported indirect chemical evidence that more than half of young, hot WDs ($T_{\rm eff} \gtrsim 20\,000 \, {\rm K}$) retain giant planets on scales of $\sim 10$--$100 \AU$. 
There is some hope for direct detections of these planets (or, at least, robust upper limits on their occurrence rate) in years to come. 
For example, direct imaging of planets around nearby WDs is feasible because at infrared wavelengths the host stars are relatively faint and the planets are self-luminous. 
Indeed, an attempt has been made to image giant planets orbiting WDs in the Hyades cluster \citep{BZT2021}, and another direct imaging search for planets around WDs is scheduled for {\it JWST} Cycle 1 \citep{MullallyJWST}. 
In Section \ref{s:discuss:HEM:ptesimals}, we identify two individual WDs that may be good candidates for a direct imaging search in the near term.

It may be possible to detect long-period planets around older and more distant WDs via gravitational microlensing \citep[e.g.,][]{Gould+2010,Suzuki+2016}. 
The recent report of a Jupiter analogue associated with a presumed WD discovered via microlensing \citep{Blackman+2021} supports this idea. 
The {\it Nancy Grace Roman Space Telescope} \citep{Spergel+2015} will be capable of detecting planets $\gtrsim 10 \ME$ at orbital distances as large as $\sim 100 \AU$ \citep{Penny+2019}. 
To our knowledge, no published study to date has specifically estimated the number of planets orbiting WDs that might be detected by {\it Roman} (or another major microlensing survey), perhaps because solitary WDs constitute a small fraction of lens stars overall. 
However, detailed models of lensing rates toward the Galactic bulge suggest that they account for about half of lenses in the mass range $0.5$--$0.7 \MSol$ \citep[see e.g.\ Fig.\ 3 of][]{Gaudi2012}, where most polluted WDs reside. 
Thus, it is plausible that {\it Roman} will detect an appreciable number of WDs with surviving planets. 
Further study may be required to clarify {\it Roman}'s ability to detect long-period planets orbiting WDs in practice. 
For completeness, we note that gravitational lensing by nearby ($\lesssim 100 \, {\rm pc}$) objects, or `mesolensing' \citep{DiStefano2008a,DiStefano2008b}, has also been suggested as way to detect planets around known WDs \citep{Harding+2018}.

\subsubsection{Fraction of WDs with pollution} \label{s:disc:WDpol:frac}

We can gauge the occurrence rate of long-period large planets around polluted WDs based on the occurrence of giant planets around A- and F-type main-sequence stars. 
``Long-period'' for the purposes of this discussion means $a \gtrsim 3 \AU$, where planets are unlikely to be engulfed during late-stage stellar evolution \citep[e.g.,][]{MV2012,Ronco+2020}. 
Around FGK stars, for which occurrence rates of long-period planets are best constrained, radial-velocity surveys find that planets with masses $\gtrsim 1 \MJ$ and orbits between $3$ and $10 \AU$ occur at a rate of $\sim 5$ per cent \citep{Cumming+2008,Fernandes+2019,Fulton+2021}. 
At least half of these planets occur in multi-planet systems \citep{Bryan+2016}, where secular chaos is possible. 
Additionally, planets with masses in the range of $\sim 0.1$--$1 \MJ$ are several times more abundant than those with masses $\gtrsim 1 \MJ$ in the same range of orbital periods: this is supported by analyses of microlensing events \citep{Suzuki+2016}, non-repeating transits observed by {\it Kepler} \citep{FM+2016,HZW2019}, and long-term radial-velocity surveys \citep{Fulton+2021}.

The occurrence rate of long-period giant planets orbiting early-type stars, which produce most observed WDs, could differ from that around late-type stars. 
Long-term radial-velocity surveys have found evidence that the fraction of stars hosting Jovian planets increases with stellar mass \citep{Johnson+2010,Reffert+2015,Jones+2016,GMJ2018}. 
The precise nature of this correlation is the subject of ongoing debate, due to the difficulty of estimating the masses of evolved stars \citep[e.g.,][]{Lloyd2011,Malla+2020}. 
Even so, the chemical signatures of evaporating giant planets orbiting young WDs reported by \citet{Schreiber+2019} suggest that occurrence rates of $50$ per cent or greater may be warranted.

We can estimate the fraction $f_{\rm sc}$ of systems in which secular chaos produces WD pollution as follows:
\begin{equation} \label{eq:fsc_Drake}
    f_{\rm sc} = f_{\rm p} f_{\rm multi} f_{\rm belt} ,
\end{equation}
where $f_{\rm p}$ is the fraction of progenitor stars with cold large planets (as defined above), $f_{\rm multi}$ is the fraction of those planets that occur in widely-spaced pairs (or higher multiples), and $f_{\rm belt}$ is the fraction of those systems with an inner planetesimal belt that overlaps with the secular chaotic zone and thus is the source of pollution. 
The studies referred to above suggest values of $f_{\rm p} \approx 0.25$--$0.5$ and $f_{\rm multi} \gtrsim 0.5$. The factor $f_{\rm belt}$ is highly uncertain. 
In our scenario, the planetesimal belt is analogous to the Solar System's main asteroid belt, and true extrasolar analogues of this belt cannot currently be detected directly. 
Assuming that inner planetesimal belts are common alongside cold giant planets, i.e.\ $f_{\rm belt} \simeq 1$, we find
\begin{equation} \label{eq:fsc_Drake_numbers}
    f_{\rm sc} = 0.25 \left( \frac{f_{\rm p}}{0.5} \right) \left( \frac{f_{\rm multi}}{0.5} \right)
\end{equation}
The observed fraction of solitary WDs with atmospheric pollution is between 25 and 50 per cent \citep{Zuckerman+2010,KGF2014,Wilson+2019}. 
Thus, secular chaos driven by large outer planets can account for a large fraction of polluted WDs. 
We note that \citet{PM2017} have previously estimated that up to 25 per cent of all polluted WDs could be produced through the Lidov--Kozai effect driven by stellar binary companions.

\subsection{Pre-WD dynamical evolution} \label{s:disc:preWD}

When a star evolves through the AGB phase to become a WD, several effects alter the secular dynamics of its planetary system:
\begin{itemize}
    \item Bodies orbiting within a critical initial distance $a_{\rm cr}$ are engulfed by the star and presumably destroyed; the value of $a_{\rm cr}$ depends on both initial stellar mass and planetary mass.
    \item Bodies orbiting beyond $a_{\rm cr}$ experience orbital expansion. 
    If the stellar mass loss is adiabatic and isotropic, then orbital angular momentum is conserved and the initial and final semi-major axes are related by
    \begin{equation}
        a_{\rm fin} = \frac{a_{\rm init}}{y},
    \end{equation}
    where $y$ is the ratio of the WD's mass to its progenitor's mass; $1/y$ is typically between 2 and 4. All other orbital elements are conserved.
    \item Small bodies (with sizes $\lesssim 10^{3} \, {\rm km}$) experience various non-gravitational forces, including radiative forces from the enhanced stellar luminosity and drag forces from the stellar wind, that can significantly alter their orbits \citep{Dong+2010,VEG2015,VHI2019} and physical properties \citep{VJG2014,VS2020}. 
    We neglect these effects for simplicity but discuss their possible implications for our scenario in Section \ref{s:disc:preWD:late}.
\end{itemize}
During adiabatic post-MS stellar mass loss, all secular frequencies change in proportion to $1/a$ (Eqs.\ \ref{eq:om_jk}--\ref{eq:nu_jk}). 
The relative locations of secular resonances and chaotic zones are also unchanged. 
Thus, a planetesimal orbiting in the chaotic zone during the host's WD stage would have also been in the chaotic zone during the MS stage. 
The chaotic diffusion time would have been a factor of $\sim 2$--$4$ shorter in the MS stage than the WD stage.

The MS stellar lifetime can be estimated (using the upper MS mass--luminosity relation of \citealt{DK1991}) as
\begin{equation} \label{eq:tMS_scaling}
    t_{\rm MS} \approx 10 \Gyr \left( \frac{M_{\rm MS}}{\MSol} \right)^{-2.92},
\end{equation}
which gives $\sim 3 \Gyr$ for a $1.5 \MSol$ progenitor and $\sim 0.4 \Gyr$ for $3.0 \MSol$. 
There is ample time to begin depleting a planetesimal belt through secular chaos during the MS stage, judging by the results of our numerical experiments. 
Na\"{i}vely, then, one would not expect a large planetesimal population in the chaotic zone at the beginning of the WD phase. 
However, there are plausible ways to circumvent this issue and allow for a planetesimal reservoir to persist around the WD. 
We discuss three possibilities in the remainder of this section.

\subsubsection{Dynamical influence of inner planets} \label{s:disc:preWD:inner}

The perturbers in our scenario are large, distant planets. 
However, observations show that giant planets are rarely the only members of a system. 
Analyses of the transiting and radial-velocity exoplanet samples have found that nearly all long-period Jovian planets coexist with inner systems of super-Earths or sub-Neptunes \citep{ZW2018,Bryan+2019}, at least for FGK host stars. 
\citet{HZW2019} have suggested that this trend also holds for long-period Neptune-like planets.

We show here that the dynamical influence of these inner planets can prevent extreme eccentricity growth of planetesimals during the MS phase, thereby preserving a reservoir of unstable material to pollute the WD. 
An inner planetary system changes the free apsidal precession rate of a coplanar test particle through its combined quadrupole potential \citep[e.g.,][]{MD1999,VC2018}. 
This contribution is
\begin{equation} \label{eq:def_omega_in}
    \omega_{\rm in} \simeq \frac{3}{4} \frac{K_{\rm in}}{M_{*} a_{\rm p}^{2}} \frac{n_{\rm p}}{(1 - e_{\rm p}^{2})^{2}},
\end{equation}
where
\begin{equation} \label{eq:def_Kin}
    K_{\rm in} \equiv \sum_{a_{j} < a_{\rm p}} m_{j} a_{j}^{2}.
\end{equation}
For simplicity, we have assumed that $a_{j} \ll a_{\rm p}$ and that the inner planets have circular orbits that lie in the invariable plane. 
The combined quadrupole potential of the inner system acts as an effective SRF greatly exceeding the GR correction, thus modifying the limiting eccentricity. Similarly to Eq.\ (\ref{eq:elim_eps1PN}), we find
\begin{equation} \label{eq:elim_inner}
    1 - e_{\rm lim} \sim \left( \frac{K_{\rm in}}{a_{\rm p}^{5} H_{\rm out}} \right)^{2/3},
\end{equation}
where
\begin{equation} \label{eq:def_Hout}
    H_{\rm out} \equiv \sum_{a_{k}>a_{\rm p}} \frac{m_{k}}{a_{k}^{3}}
\end{equation}
measures the strength of the outer planets' quadrupole potential.

For a specific example, let us consider the planetary system of experiment N0 (see Table \ref{tab:N0_params}) as it would have been during the host's MS phase. 
Using the initial--final mass relation of \citet{Cummings+2018}, we estimate that the progenitor our fiducial WD ($0.6 \MSol$) had a mass of $1.5 \MSol$, for a mass loss factor $y = 0.6/1.5 = 0.4$. 
Thus, the perturbers would have been located at $a_{1,{\rm MS}} = 8.0 \AU$ and $a_{2,{\rm MS}} = 17.2 \AU$, and the midpoint of the planetesimal belt would have been at $a_{\rm MS} = 2.4 \AU$ (assuming orbital expansion under adiabatic stellar mass loss only, but see Section \ref{s:disc:preWD:late}). 
The outer planets' quadrupole strength during the MS is $H_{\rm out} = 0.0625 \ME \AU^{-3}$. 
According to Eq.\ (\ref{eq:elim_inner}), particles at $a = 2.4 \AU$ would have their limiting eccentricities reduced to $e_{\rm lim} \lesssim 0.5$ if an inner system with $K_{\rm in} \gtrsim 1.75 \ME \AU^{2}$ were present. 
This can be satisfied by a system of super-Earth- to Neptune-sized planets extending to $\sim 0.5$--$1 \AU$ or by a warm Jupiter at $\sim 0.1 \AU$.

We have verified that the above is qualitatively accurate by performing an additional set of numerical experiments using the `MS N0' setup described above. 
That is, we use the standard initial condition distributions found in Tables \ref{tab:N0_params} and \ref{tab:tpchaos_ICs} except with $M_{*} = 1.5 \MSol$ and orbits reduced in size by a factor of $0.4$. 
We include in {\sc rings} an additional planet on a circular orbit aligned with the invariable plane with semi-major axis $a_{\rm in} \ll a_{\rm p}$ and mass $m_{\rm in}$ such that $K_{\rm in} = m_{\rm in} a_{\rm in}^{2}$. 
We conduct 100 runs for each value of $K_{\rm in}$, each lasting up to $3 \Gyr$ (per Eq.\ \ref{eq:tMS_scaling}). 
We terminate a run if the particle's eccentricity exceeds $0.9$, to prevent orbit crossing. 
With $K_{\rm in} = 0$, we find a degree of chaos similar to the original N0, as expected: a majority of test particles (62/100) are excited to eccentricity greater than $0.9$. 
The results are similar again (68/100) for $K_{\rm in} = 0.3 \ME \AU^{2}$. For $K_{\rm in} = 3 \ME \AU^{2}$, however, we find that the prevalence of chaotic diffusion is greatly reduced: only 17/100 particles exceed an eccentricity of $0.9$, while the rest never reach $0.5$.

The engulfment of the inner planets during the host star's AGB phase removes this effective SRF and approximately coincides with the orbital expansion of the planetesimal belt and outer planets. 
Thus, tidal disruption events driven by secular chaos can be naturally delayed until the WD stage \citep[see also][]{PM2017,Smallwood+2018,Smallwood+2021}.

\subsubsection{Late redistribution of planetesimals under non-gravitational forces} \label{s:disc:preWD:late}

As mentioned earlier in this section, the dynamical evolution of small bodies during late-stage stellar evolution can be nontrivial due to the effects of non-gravitational forces \citep[e.g.,][]{Dong+2010,VJG2014,VEG2015,VHI2019}. 
These influences are highly sensitive to a body's size and are negligible for bodies larger than $\sim 10^{3} \, {\rm km}$. 
Bodies smaller than this limit can migrate differentially with respect to major planets, either inward and outward, potentially bringing them into (or out of) chaotic zones.

A detailed calculation of how much planetesimal mass could be placed in a chaotic zone by post-MS non-gravitational forces is beyond the scope of this study. 
It is conceivable, however, that the chaotic zone could be replenished in this manner just before the WD phase. 
If planetesimal belts around MS stars are characteristically much more massive than the Solar System main belt, then only a small fraction of its original mass needs to be inserted into the chaotic zone after stellar evolution in order to account for the observed metal accretion rates of WDs (Fig.\ \ref{fig:Mext_age}).

Non-gravitational forces during late-stage stellar evolution also influence the mass distribution of surviving planetesimals. 
For example, radiative torques associated with the YORP effect can spin up rubble-pile asteroids to their breakup rate, destroying all objects with radii less than $\sim 10 \, {\rm km}$ within $\sim 10 \AU$ of an AGB star \citep{VJG2014}. 
This effect alone this does not reduce the amount of {\it mass} available to pollute the WD. 
However, it does affect the {\it number} of available objects, which in turn affects the observed metal accretion rates of WDs as a function of cooling age by changing the average mass delivered per accretion event and the average interval between events (see Section \ref{s:discuss:Mdot_vs_age} and \citealt{Wyatt+2014}). Suppose that (i) the present-day Solar System main belt is representative of extrasolar planetesimal belts in terms of total population and size distribution; and (ii) the only non-gravitational effect is the YORP-induced destruction of all objects with radius less than $10 \, {\rm km}$. 
We can estimate the number of surviving objects by querying the on-line Jet Propulsion Laboratory Small-Body Database\footnote{\url{https://ssd.jpl.nasa.gov/tools/sbdb_query.html}. 
Accessed 2022 February 20.} for all main-belt asteroids with a listed {\it diameter} greater than $20 \, {\rm km}$: this returns $1733$ objects. 
This suggests that the nominal value $N_{0} = 10^{3}$ used in Eq.\ (\ref{eq:Delta_tdis}) is of the correct order of magnitude, given our simplistic assumptions. 
The assumed critical size for survival is moderately important: repeating the query for diameters greater than $10 \, {\rm km}$ ($30 \, {\rm km}$) returns $6936$ ($976$) objects.

\subsubsection{Self-gravity of planetesimal belt} \label{s:disc:preWD:selfgravity}

In our simulations, we assumed that gravitational interactions among the planetesimals were negligible. 
However, a sufficiently massive belt would undergo additional precession due to self-gravity, providing another way for the reservoir to resist secular chaos during the star's MS lifetime. 
Depending on whether and how much the planetesimal population is depleted or redistributed during stellar evolution (Section \ref{s:disc:preWD:late}), self-gravity may become negligible during the WD phase, rendering the belt vulnerable to secular chaos. 
An exact calculation of the true precession rate due to self-gravity using linear perturbation theory \citep{TO2016,TL2019} would be too complicated for our purposes, but a reasonable approximation can be obtained by considering the gravity of the most massive object in the belt (which can contain a large fraction of the total mass, depending on the size distribution) in the limit where the belt is narrow compared to its mean radius. 
Assuming that body (mass $m_{\rm b}$, semi-major axis $a_{\rm b}$) lies near the inner edge of the belt, the precession of other planetesimals is described by a modified version of Eq.\ (\ref{eq:def_omega_in}):
\begin{equation}
    \omega_{\rm sg} = \frac{n_{\rm p}}{4} \frac{m_{\rm b}}{M_{*}} \frac{a_{\rm b}}{a_{\rm p}} b_{3/2}^{(1)}(a_{\rm b} / a_{\rm p}), \label{eq:def_omega_sg}
\end{equation}
where $a_{\rm b} / a_{\rm p}$ is slightly less than $1$. 
Suppression of secular chaos requires that $\omega_{\rm sg}$ be greater than the free precession rate due to the outer planets (e.g., Eq.\ \ref{eq:def_om_LK}); this corresponds to
\begin{align}
    m_{\rm b} &\gtrsim \frac{H_{\rm out} a_{\rm p}^{4}}{a_{\rm b}} \left[ b_{3/2}^{(1)}(a_{\rm b} / a_{\rm p}) \right]^{-1} \nonumber \\
    &= 1.25 \ME \left( \frac{H_{\rm out}}{0.1 \ME \AU^{-3}} \right) \left( \frac{a_{\rm p}}{5 \AU} \right)^{3} \nonumber \\
    & \hspace{1.5cm} \times \left[ \frac{(a_{\rm b} / a_{\rm p}) b_{3/2}^{(1)}(a_{\rm b} / a_{\rm p})}{10} \right]^{-1}, \label{eq:self_grav_mb}
\end{align}
where $H_{\rm out}$ is the quadrupole strength of the outer planets (Eq.\ \ref{eq:def_Hout}). 
If the most massive body lies near the outer edge of the belt instead, we have
\begin{equation}
    \omega_{\rm sg} = \frac{n_{\rm p}}{4} \frac{m_{\rm b}}{M_{*}} \left( \frac{a_{\rm p}}{a_{\rm b}} \right)^{2} b_{3/2}^{(1)}(a_{\rm p} / a_{\rm b}). \label{eq:def_omega_sg_2}
\end{equation}
This leads to a result that is effectively identical to Eq.\ (\ref{eq:self_grav_mb}) because $a_{\rm p} / a_{\rm b}$ is close to unity. 
The estimated value of $m_{\rm b}$ in Eq.\ (\ref{eq:self_grav_mb}) should be viewed as a lower bound on the belt mass required to suppress chaos because a real planetesimal belt likely has significant width compared to its radius, decreasing the Laplace coefficients in Eqs.\ (\ref{eq:def_omega_sg}) and (\ref{eq:def_omega_sg_2}). 
Thus, the self-gravity of a planetesimal belt can suppress secular chaos driven by outer giant planets only if the belt is several orders of magnitude more massive than the Solar System's main asteroid belt ($\sim 5 \times 10^{-4} \ME$). 
This is possible (but not required) based on current theories of the asteroid belt's origins and dynamical evolution \citep[e.g.][]{RN2020}. 
Again, if the typical mass of extrasolar planetesimal belts around MS stars is much greater than that of the Solar System main belt, then only a small fraction of that mass needs to survive in the chaotic zone to subsequently pollute the WD. 
We note also that the initial belt mass required in order to reproduce the observed metal accretion rates of polluted WDs via secular chaos ($\sim 10^{-3} \ME$) is well below the amount at which self-gravity becomes important -- consistent with our initial assumptions.

\subsection{High-eccentricity migration of planets and planetesimals around WDs} \label{s:discuss:HEM}

\subsubsection{Circularization of planetesimals} \label{s:discuss:HEM:ptesimals}

We expect most planetesimals that are excited onto small-pericentre orbits via secular chaos to be tidally disrupted and accreted by the WD. 
However, under certain conditions, it is possible for a planetesimal to be circularized to a short-period ($< 1 \, \dif$) orbit. 
Indeed, circularized objects with periods $\sim 2$--$5 \, {\rm h}$ have been reported in the systems WD\,1145+017 \citep{Vanderburg+2015} and SDSS\,J1228+1040 \citep{Manser+2019}. 
There are at least two possible circularization mechanisms: tidal dissipation within the planetesimal \citep{OL2020} and drag forces due to interaction with a compact accretion disc around the WD \citep{GV2019,OL2020,MGB2021}. 
In light of our study of secular chaos, what can we say about the possible dynamical evolution of these objects before their circularization?

It is interesting that both WDs with circularized planetesimals are relatively young, with cooling ages of $100$--$200 \Myr$; and that both are apparently solitary, with no mention of stellar companions in the literature to our knowledge. 
This suggests that long-term planet--planet interactions such as secular chaos are responsible for the high-eccentricity migration of these objects. 
Our simulation results let us place quantitative constraints on the properties of hypothetical planetary companions driving secular chaos in these systems.

The circularized objects in both systems are currently disintegrating, with expected lifetimes $\lesssim 1 \Myr$ \citep[see e.g.\ the Supplementary Materials of][]{Manser+2019}. 
It is likely that they migrated to their observed orbits quite recently, say $\lesssim 1 \Myr$ ago. 
At the same time, the fact that so few of these circularized bodies have been discovered among $>1000$ known polluted WDs suggests that circularization may be improbable compared to disruption. 
Thus, one would expect to observe circularized bodies mostly in systems where the rate of dynamical excitation is at an all-time high. 
We suggest that the planetary architecture in these systems is such that secular chaos produces a peak in the tidal disruption rate near their present cooling ages, i.e.\ around $100$ and $200 \Myr$ for SDSS\,J1228+1040 and WD\,1145+017, respectively. 
Referring back to our numerical experiments (Section \ref{s:expts:main}), we see that experiment A2, with two $\sim$ Jupiter-mass planets at $20$ and $43 \AU$, has this property. Of course, it is possible to adjust these values to produce a peak at the desired time in a variety of ways, such as re-scaling the masses and semi-major axes of the planets (while preserving $m_{j}/a_{j}^{3/2}$; Eq.\ \ref{eq:secular_rescaling}) or adjusting the amplitudes of their secular modes. 
In any case, we can rule out all configurations that produce a peak in the disruption rate at cooling ages much greater or less than $\sim 100$--$200 \Myr$.

In Section \ref{s:discuss:WDpol:detect}, we discussed prospects for direct detection of outer planets driving secular chaos around WDs. 
We noted that young, nearby WDs are currently of interest as targets for direct imaging searches. 
The SDSS\,J1228+1040 and WD\,1145+017 systems may be good candidates for these campaigns on both counts. 
Medium-to-high-contrast imaging with an angular resolution of 100 mas could probe these systems for planets with sky-projected separations as small as 12.5 and 17.5 au, respectively, at their distances of 125 and 175 pc. SDSS\,J1228+1040 is better suited for a direct imaging search, both because it is the nearer of the two systems and because it has a total age (= MS lifetime + cooling age) less than $1 \Gyr$ based on the measured WD mass of $0.705 \MSol$ \citep{KGF2014} and estimated progenitor mass of $2.70 \MSol$ \citep{Cummings+2018}. 
Based on the near-infrared luminosity of $\sim 10^{-4} L_{\odot}$ for the WD \citep[which, in this system, is dominated by a compact accretion disc:][]{Gansicke+2006,Brinkworth+2009,Debes+2011} and an estimated $\sim 10^{-8} L_{\odot}$ for a $1 \MJ$ planet at an age of $1 \Gyr$ \citep{Baraffe+2003}, we estimate that a planet-to-star contrast of $10^{-4}$ is required for detection. 
This will soon be achievable for angular separations over $100 \, {\rm mas}$ with current ground-based facilities and perhaps at smaller separations with planned high-contrast imagers on 30-m-class telescopes \citep[e.g.,][]{Lawson+2012}. 
Despite the challenges of direct imaging, we believe the systems discussed above present a significant opportunity to investigate the simultaneous dynamical evolution of planets and planetesimals around WDs.

\subsubsection{Circularization and disruption of surviving planets} \label{s:discuss:HEM:planets}

Thus far we have interpreted the test particle in our numerical simulations as a planetesimal. 
However, our results can be extended qualitatively to the dynamical evolution of systems of three (or more) planets orbiting a WD, with the innermost planet replacing the planetesimal. 
Indeed, recent discoveries suggest that surviving planets orbiting WDs can be excited to extreme eccentricities. 
Our results may be relevant to these observations.

One system of interest is WD\,J0914+1914, where the WD is accreting matter from a disrupted or evaporating ice giant \citep{Gansicke+2019}. 
While this system has a short cooling age ($\sim 14 \Myr$), our results and those of \citet{TLV2019} indicate that planet--planet secular chaos can lead to planetary tidal disruption events within that time, given a suitable system configuration (e.g., experiment A2 in Fig.\ \ref{fig:logt_m1}). 
\citet{Stephan+2020} have shown that the Lidov--Kozai mechanism can likewise lead to tidal disruption of a giant planet in this system within its cooling age and have constrained the properties of a hypothetical binary companion in this scenario. 
We suggest that, if no such companion is found, planet--planet secular chaos may better explain the dynamical evolution of WD\,J0914+1914\,b. 
The disrupted planet would have been the innermost member of the system and would have two or more outer planetary companions that drove its chaos. 
Unfortunately, this system is at a distance of $\sim 600 \, {\rm pc}$, probably too far for direct imaging of these outer planets to be possible in the near term.

Another object of interest is WD\,1856+534\,b, an intact, Jupiter-sized body transiting its WD host with a period of $1.4 \, {\rm d}$ \citep{Vanderburg+2020}. 
The mass of the transiting companion is not known precisely but is currently constrained within the range of $\sim 1$--$14 \MJ$ \citep{Vanderburg+2020,Xu+2021}. 
Various dynamical histories have been proposed for this system, including high-eccentricity migration through the Lidov--Kozai mechanism driven by the WD's bound stellar companions \citep*{MP2020,OConnor+2020,Stephan+2020}. 
While WD\,1856+534\,b is currently the only known planet of its kind, others may be discovered by future surveys. 
If a significant number of these planets are found to orbit solitary WDs, planet--planet secular chaos could be considered as an explanation for their migration.

\section{Conclusion} \label{s:conclude}

In this article, we have explored the occurrence of secular chaos in remnant planetary systems orbiting WDs. 
This process can drive tidal disruption of planetesimals over time-scales ranging from a few Myr to many Gyr, thereby producing a steady rate of atmospheric metal pollution across all WD cooling ages. 
Secular chaos may also cause high-eccentricity migration of surviving planets and planetesimals around WDs, producing systems like WD\,J0914+1914, WD\,1856+534, WD\,1145+017, and SDSS\,J1228+1040. 
Upcoming surveys may discover a significant number of both short- and long-period planets around WDs, putting our proposed mechanism to the test.

We summarize our main results below:
\begin{enumerate}
    \item[(i)] Secular chaos occurs readily in planetary systems with widely-space orbits and moderate eccentricities and mutual inclinations. 
    A necessary condition to pollute a WD via secular chaos is that a reservoir of planetesimals be located in a region where non-linear secular resonances overlap. 
    The system's orbits and masses can be re-scaled to produce the same secular behaviour for the same initial conditions in eccentricity, inclination, and orbital angles (Eq.\ \ref{eq:secular_rescaling}).
    \item[(ii)] Our numerical simulations show that when secular chaos acts on a large number of planetesimals, it produces tidal disruption events steadily over a long time span. 
    The disruption rate is well approximated by a log-normal distribution at early times and a weak declining power law at late times (Eq.\ \ref{eq:TDE_rate_lognormal_1}). 
    The time of the maximum disruption rate is inversely proportional to the masses and eccentricities of the outer planets that drive the chaos but depends only weakly on the planetesimal's exact location within the chaotic zone.
    \item[(iii)] Adopting several fiducial planetary system configurations, we constrain the required mass of a remnant planetesimal belt to reproduce the observed metal accretion rates of WDs. 
    Observations suggest that WD pollution begins around a cooling age of $\sim 30$--$50 \Myr$, that the typical external supply rate of metals is $\sim 10^{7}$--$10^{9} \, {\rm g \, s^{-1}}$ for WDs with cooling ages $\lesssim 1 \Gyr$, and that the supply declines gradually for ages $\gtrsim 1 \Gyr$ (Fig.\ \ref{fig:Mext_age}). 
    Many of our simulations of secular chaos reproduce all of these trends if the initial mass of the remnant belt in the chaotic zone is $\sim 10^{-3} \ME$, similar to the present-day main asteroid belt in the Solar System.
    \item[(iv)] Our calculations suggest that a large fraction of WDs could be polluted via secular chaos (Section \ref{s:disc:WDpol:frac}, Eq.\ \ref{eq:fsc_Drake_numbers}). 
    The actual percentage depends on the occurrence of long-period giant planets and inner planetesimal belts around WDs and, by extension, F- and A-type stars.
    \item[(v)] The secular chaos mechanism requires that a polluted WD host at least two planets with masses $\gtrsim 10 \ME$ and semi-major axes $\gtrsim 10 \AU$. 
    These planets may be detectable through direct imaging, if they have masses $\gtrsim 1 \MJ$ and the host system is relatively young and nearby; or microlensing, if the system is distant.
    \item[(vi)] We consider the possibility of circularizing planetesimals and planets on short-period orbits around WDs after their orbits are excited through secular chaos. 
    We can constrain the possible outer-planet architectures of the two WD systems with known short-period planetesimals, WD\,1145+017 and SDSS\,J1228+1040, based on the argument that circularized bodies are most likely to be observed when the system's dynamical excitation rate peaks. 
    SDSS\,J1228+1040 is nearby ($125 \, {\rm pc}$) and relatively young (cooling age $\sim 100 \Myr$, total age $\lesssim 1 \Gyr$), suggesting that direct imaging of its hypothetical outer planets may be possible in the near term.
\end{enumerate}

Finally, we reiterate some caveats of our study. 
We have assumed that long-period planetary systems are usually accompanied by an inner planetesimal belt in the chaotic zone determined by overlapping secular resonances. 
If the actual occurrence rate of these belts is small, then the expected fraction of systems in which secular chaos produces pollution would be significantly less than our nominal estimate. 
To resolve this uncertainty, a more complete census of extrasolar planetesimal populations is desirable, particularly in systems with long-period giant planets. 
Additionally, we have not simulated the evolution of the system over the full stellar lifetime. 
Pre-WD dynamical evolution is potentially important, both because secular chaos can deplete a planetesimal belt within the host star's MS lifetime and because planetesimals are expected to evolve differently from planets during the late stages of stellar evolution. 
Future studies could address these self-consistently by including secular perturbations from an inner planetary system during the MS stage and implementing a prescription for the radial migration of planets and planetesimals during post-MS evolution within a ring-averaging secular code.

\section*{Acknowledgements}

We are grateful to Will M.\ Farr for making the code {\sc rings} publicly available and to Laetitia Rodet for helpful discussions about direct imaging. 
We also thank the referee for a thorough review; their suggestions greatly improved the manuscript. 
This work has been supported in part by National Science Foundation grant AST-2107796 and NASA grant 80NSSC19K0444. 
C.E.O.\ is supported in part by a Space Grant Graduate Fellowship from the NASA New York Space Grant Consortium and by a Sadov Graduate Student Fellowship from Cornell University. 
J.T.\ is supported by a Fonds de la Recherche Scientifique -- FNRS Postdoctoral Research Fellowship. 
This research has made use of NASA's Astrophysics Data System and of the software libraries {\sc matplotlib} \citep{Hunter2007}, {\sc numpy} \citep{Harris+2020}, and {\sc scipy} \citep{Virtanen+2020}.

\section*{Data availability}

The data underlying this article will be shared on reasonable request to the corresponding author.

\bibliography{mn_21_4272_final}

\begin{thebibliography}{}
\makeatletter
\relax
\def\mn@urlcharsother{\let\do\@makeother \do\$\do\&\do\#\do\^\do\_\do\%\do\~}
\def\mn@doi{\begingroup\mn@urlcharsother \@ifnextchar [ {\mn@doi@}
  {\mn@doi@[]}}
\def\mn@doi@[#1]#2{\def\@tempa{#1}\ifx\@tempa\@empty \href
  {http://dx.doi.org/#2} {doi:#2}\else \href {http://dx.doi.org/#2} {#1}\fi
  \endgroup}
\def\mn@eprint#1#2{\mn@eprint@#1:#2::\@nil}
\def\mn@eprint@arXiv#1{\href {http://arxiv.org/abs/#1} {{\tt arXiv:#1}}}
\def\mn@eprint@dblp#1{\href {http://dblp.uni-trier.de/rec/bibtex/#1.xml}
  {dblp:#1}}
\def\mn@eprint@#1:#2:#3:#4\@nil{\def\@tempa {#1}\def\@tempb {#2}\def\@tempc
  {#3}\ifx \@tempc \@empty \let \@tempc \@tempb \let \@tempb \@tempa \fi \ifx
  \@tempb \@empty \def\@tempb {arXiv}\fi \@ifundefined
  {mn@eprint@\@tempb}{\@tempb:\@tempc}{\expandafter \expandafter \csname
  mn@eprint@\@tempb\endcsname \expandafter{\@tempc}}}

\bibitem[\protect\citeauthoryear{{Baraffe}, {Chabrier}, {Barman}, {Allard}  \&
  {Hauschildt}}{{Baraffe} et~al.}{2003}]{Baraffe+2003}
{Baraffe} I.,  {Chabrier} G.,  {Barman} T.~S.,  {Allard} F.,   {Hauschildt}
  P.~H.,  2003, \mn@doi [\aap] {10.1051/0004-6361:20030252}, \href
  {https://ui.adsabs.harvard.edu/abs/2003A&A...402..701B} {402, 701}

\bibitem[\protect\citeauthoryear{{Blackman} et~al.,}{{Blackman}
  et~al.}{2021}]{Blackman+2021}
{Blackman} J.~W.,  et~al., 2021, \mn@doi [\nat] {10.1038/s41586-021-03869-6},
  \href {https://ui.adsabs.harvard.edu/abs/2021Natur.598..272B} {598, 272}

\bibitem[\protect\citeauthoryear{{Blouin} \& {Xu}}{{Blouin} \&
  {Xu}}{2022}]{BX2022}
{Blouin} S.,  {Xu} S.,  2022, \mn@doi [\mnras] {10.1093/mnras/stab3446}, \href
  {https://ui.adsabs.harvard.edu/abs/2022MNRAS.510.1059B} {510, 1059}

\bibitem[\protect\citeauthoryear{{Bonsor} \& {Wyatt}}{{Bonsor} \&
  {Wyatt}}{2010}]{BW2010}
{Bonsor} A.,  {Wyatt} M.,  2010, \mn@doi [\mnras]
  {10.1111/j.1365-2966.2010.17412.x}, \href
  {https://ui.adsabs.harvard.edu/abs/2010MNRAS.409.1631B} {409, 1631}

\bibitem[\protect\citeauthoryear{{Bonsor}, {Mustill}  \& {Wyatt}}{{Bonsor}
  et~al.}{2011}]{BMW2011}
{Bonsor} A.,  {Mustill} A.~J.,   {Wyatt} M.~C.,  2011, \mn@doi [\mnras]
  {10.1111/j.1365-2966.2011.18524.x}, \href
  {https://ui.adsabs.harvard.edu/abs/2011MNRAS.414..930B} {414, 930}

\bibitem[\protect\citeauthoryear{{Bonsor}, {Carter}, {Hollands},
  {G{\"a}nsicke}, {Leinhardt}  \& {Harrison}}{{Bonsor}
  et~al.}{2020}]{Bonsor+2020}
{Bonsor} A.,  {Carter} P.~J.,  {Hollands} M.,  {G{\"a}nsicke} B.~T.,
  {Leinhardt} Z.,   {Harrison} J. H.~D.,  2020, \mn@doi [\mnras]
  {10.1093/mnras/stz3603}, \href
  {https://ui.adsabs.harvard.edu/abs/2020MNRAS.492.2683B} {492, 2683}

\bibitem[\protect\citeauthoryear{{Brandner}, {Zinnecker}  \&
  {Kopytova}}{{Brandner} et~al.}{2021}]{BZT2021}
{Brandner} W.,  {Zinnecker} H.,   {Kopytova} T.,  2021, \mn@doi [\mnras]
  {10.1093/mnras/staa3422}, \href
  {https://ui.adsabs.harvard.edu/abs/2021MNRAS.500.3920B} {500, 3920}

\bibitem[\protect\citeauthoryear{{Brinkworth}, {G{\"a}nsicke}, {Marsh}, {Hoard}
   \& {Tappert}}{{Brinkworth} et~al.}{2009}]{Brinkworth+2009}
{Brinkworth} C.~S.,  {G{\"a}nsicke} B.~T.,  {Marsh} T.~R.,  {Hoard} D.~W.,
  {Tappert} C.,  2009, \mn@doi [\apj] {10.1088/0004-637X/696/2/1402}, \href
  {https://ui.adsabs.harvard.edu/abs/2009ApJ...696.1402B} {696, 1402}

\bibitem[\protect\citeauthoryear{{Bryan} et~al.,}{{Bryan}
  et~al.}{2016}]{Bryan+2016}
{Bryan} M.~L.,  et~al., 2016, \mn@doi [\apj] {10.3847/0004-637X/821/2/89},
  \href {https://ui.adsabs.harvard.edu/abs/2016ApJ...821...89B} {821, 89}

\bibitem[\protect\citeauthoryear{{Bryan}, {Knutson}, {Lee}, {Fulton},
  {Batygin}, {Ngo}  \& {Meshkat}}{{Bryan} et~al.}{2019}]{Bryan+2019}
{Bryan} M.~L.,  {Knutson} H.~A.,  {Lee} E.~J.,  {Fulton} B.~J.,  {Batygin} K.,
  {Ngo} H.,   {Meshkat} T.,  2019, \mn@doi [\aj] {10.3847/1538-3881/aaf57f},
  \href {https://ui.adsabs.harvard.edu/abs/2019AJ....157...52B} {157, 52}

\bibitem[\protect\citeauthoryear{{Chen} et~al.,}{{Chen}
  et~al.}{2019}]{Chen+2019}
{Chen} D.-C.,  et~al., 2019, \mn@doi [Nature Astronomy]
  {10.1038/s41550-018-0609-7}, \href
  {https://ui.adsabs.harvard.edu/abs/2019NatAs...3...69C} {3, 69}

\bibitem[\protect\citeauthoryear{{Chirikov}}{{Chirikov}}{1979}]{Chirikov1979}
{Chirikov} B.~V.,  1979, \mn@doi [\physrep] {10.1016/0370-1573(79)90023-1},
  \href {https://ui.adsabs.harvard.edu/abs/1979PhR....52..263C} {52, 263}

\bibitem[\protect\citeauthoryear{{Cumming}, {Butler}, {Marcy}, {Vogt}, {Wright}
   \& {Fischer}}{{Cumming} et~al.}{2008}]{Cumming+2008}
{Cumming} A.,  {Butler} R.~P.,  {Marcy} G.~W.,  {Vogt} S.~S.,  {Wright} J.~T.,
   {Fischer} D.~A.,  2008, \mn@doi [\pasp] {10.1086/588487}, \href
  {https://ui.adsabs.harvard.edu/abs/2008PASP..120..531C} {120, 531}

\bibitem[\protect\citeauthoryear{{Cummings}, {Kalirai}, {Tremblay},
  {Ramirez-Ruiz}  \& {Choi}}{{Cummings} et~al.}{2018}]{Cummings+2018}
{Cummings} J.~D.,  {Kalirai} J.~S.,  {Tremblay} P.~E.,  {Ramirez-Ruiz} E.,
  {Choi} J.,  2018, \mn@doi [\apj] {10.3847/1538-4357/aadfd6}, \href
  {https://ui.adsabs.harvard.edu/abs/2018ApJ...866...21C} {866, 21}

\bibitem[\protect\citeauthoryear{{Davidsson}}{{Davidsson}}{1999}]{Davidsson1999}
{Davidsson} B. J.~R.,  1999, \mn@doi [\icarus] {10.1006/icar.1999.6214}, \href
  {https://ui.adsabs.harvard.edu/abs/1999Icar..142..525D} {142, 525}

\bibitem[\protect\citeauthoryear{{Debes} \& {Sigurdsson}}{{Debes} \&
  {Sigurdsson}}{2002}]{DS2002}
{Debes} J.~H.,  {Sigurdsson} S.,  2002, \mn@doi [\apj] {10.1086/340291}, \href
  {https://ui.adsabs.harvard.edu/abs/2002ApJ...572..556D} {572, 556}

\bibitem[\protect\citeauthoryear{{Debes}, {Hoard}, {Wachter}, {Leisawitz}  \&
  {Cohen}}{{Debes} et~al.}{2011}]{Debes+2011}
{Debes} J.~H.,  {Hoard} D.~W.,  {Wachter} S.,  {Leisawitz} D.~T.,   {Cohen} M.,
   2011, \mn@doi [\apjs] {10.1088/0067-0049/197/2/38}, \href
  {https://ui.adsabs.harvard.edu/abs/2011ApJS..197...38D} {197, 38}

\bibitem[\protect\citeauthoryear{{Debes}, {Walsh}  \& {Stark}}{{Debes}
  et~al.}{2012}]{DWS2012}
{Debes} J.~H.,  {Walsh} K.~J.,   {Stark} C.,  2012, \mn@doi [\apj]
  {10.1088/0004-637X/747/2/148}, \href
  {https://ui.adsabs.harvard.edu/abs/2012ApJ...747..148D} {747, 148}

\bibitem[\protect\citeauthoryear{{Demircan} \& {Kahraman}}{{Demircan} \&
  {Kahraman}}{1991}]{DK1991}
{Demircan} O.,  {Kahraman} G.,  1991, \mn@doi [\apss] {10.1007/BF00639097},
  \href {https://ui.adsabs.harvard.edu/abs/1991Ap&SS.181..313D} {181, 313}

\bibitem[\protect\citeauthoryear{{Di Stefano}}{{Di
  Stefano}}{2008a}]{DiStefano2008a}
{Di Stefano} R.,  2008a, \mn@doi [\apj] {10.1086/524395}, \href
  {https://ui.adsabs.harvard.edu/abs/2008ApJ...684...46D} {684, 46}

\bibitem[\protect\citeauthoryear{{Di Stefano}}{{Di
  Stefano}}{2008b}]{DiStefano2008b}
{Di Stefano} R.,  2008b, \mn@doi [\apj] {10.1086/528940}, \href
  {https://ui.adsabs.harvard.edu/abs/2008ApJ...684...59D} {684, 59}

\bibitem[\protect\citeauthoryear{{Dong}, {Wang}, {Lin}  \& {Liu}}{{Dong}
  et~al.}{2010}]{Dong+2010}
{Dong} R.,  {Wang} Y.,  {Lin} D.~N.~C.,   {Liu} X.~W.,  2010, \mn@doi [\apj]
  {10.1088/0004-637X/715/2/1036}, \href
  {https://ui.adsabs.harvard.edu/abs/2010ApJ...715.1036D} {715, 1036}

\bibitem[\protect\citeauthoryear{{Doyle}, {Desch}  \& {Young}}{{Doyle}
  et~al.}{2021}]{DDY2021}
{Doyle} A.~E.,  {Desch} S.~J.,   {Young} E.~D.,  2021, \mn@doi [\apjl]
  {10.3847/2041-8213/abd9ba}, \href
  {https://ui.adsabs.harvard.edu/abs/2021arXiv210201835D} {907, L35}

\bibitem[\protect\citeauthoryear{{Fabrycky} \& {Tremaine}}{{Fabrycky} \&
  {Tremaine}}{2007}]{FT2007}
{Fabrycky} D.,  {Tremaine} S.,  2007, \mn@doi [\apj] {10.1086/521702}, \href
  {https://ui.adsabs.harvard.edu/abs/2007ApJ...669.1298F} {669, 1298}

\bibitem[\protect\citeauthoryear{{Farihi}, {Jura}  \& {Zuckerman}}{{Farihi}
  et~al.}{2009}]{FJZ2009}
{Farihi} J.,  {Jura} M.,   {Zuckerman} B.,  2009, \mn@doi [\apj]
  {10.1088/0004-637X/694/2/805}, \href
  {https://ui.adsabs.harvard.edu/abs/2009ApJ...694..805F} {694, 805}

\bibitem[\protect\citeauthoryear{{Farihi} et~al.,}{{Farihi}
  et~al.}{2022}]{Farihi+2022}
{Farihi} J.,  et~al., 2022, \mn@doi [\mnras] {10.1093/mnras/stab3475}, \href
  {https://ui.adsabs.harvard.edu/abs/2022MNRAS.511.1647F} {511, 1647}

\bibitem[\protect\citeauthoryear{{Fernandes}, {Mulders}, {Pascucci},
  {Mordasini}  \& {Emsenhuber}}{{Fernandes} et~al.}{2019}]{Fernandes+2019}
{Fernandes} R.~B.,  {Mulders} G.~D.,  {Pascucci} I.,  {Mordasini} C.,
  {Emsenhuber} A.,  2019, \mn@doi [\apj] {10.3847/1538-4357/ab0300}, \href
  {https://ui.adsabs.harvard.edu/abs/2019ApJ...874...81F} {874, 81}

\bibitem[\protect\citeauthoryear{{Foreman-Mackey}, {Morton}, {Hogg}, {Agol}  \&
  {Sch{\"o}lkopf}}{{Foreman-Mackey} et~al.}{2016}]{FM+2016}
{Foreman-Mackey} D.,  {Morton} T.~D.,  {Hogg} D.~W.,  {Agol} E.,
  {Sch{\"o}lkopf} B.,  2016, \mn@doi [\aj] {10.3847/0004-6256/152/6/206}, \href
  {https://ui.adsabs.harvard.edu/abs/2016AJ....152..206F} {152, 206}

\bibitem[\protect\citeauthoryear{{Frewen} \& {Hansen}}{{Frewen} \&
  {Hansen}}{2014}]{FH2014}
{Frewen} S.~F.~N.,  {Hansen} B.~M.~S.,  2014, \mn@doi [\mnras]
  {10.1093/mnras/stu097}, \href
  {https://ui.adsabs.harvard.edu/abs/2014MNRAS.439.2442F} {439, 2442}

\bibitem[\protect\citeauthoryear{{Fulton} et~al.,}{{Fulton}
  et~al.}{2021}]{Fulton+2021}
{Fulton} B.~J.,  et~al., 2021, \mn@doi [\apjs] {10.3847/1538-4365/abfcc1},
  \href {https://ui.adsabs.harvard.edu/abs/2021ApJS..255...14F} {255, 14}

\bibitem[\protect\citeauthoryear{{G{\"a}nsicke}, {Marsh}, {Southworth}  \&
  {Rebassa-Mansergas}}{{G{\"a}nsicke} et~al.}{2006}]{Gansicke+2006}
{G{\"a}nsicke} B.~T.,  {Marsh} T.~R.,  {Southworth} J.,   {Rebassa-Mansergas}
  A.,  2006, \mn@doi [Science] {10.1126/science.1135033}, \href
  {https://ui.adsabs.harvard.edu/abs/2006Sci...314.1908G} {314, 1908}

\bibitem[\protect\citeauthoryear{{G{\"a}nsicke}, {Schreiber}, {Toloza},
  {Gentile Fusillo}, {Koester}  \& {Manser}}{{G{\"a}nsicke}
  et~al.}{2019}]{Gansicke+2019}
{G{\"a}nsicke} B.~T.,  {Schreiber} M.~R.,  {Toloza} O.,  {Gentile Fusillo}
  N.~P.,  {Koester} D.,   {Manser} C.~J.,  2019, \mn@doi [\nat]
  {10.1038/s41586-019-1789-8}, \href
  {https://ui.adsabs.harvard.edu/abs/2019Natur.576...61G} {576, 61}

\bibitem[\protect\citeauthoryear{{Gaudi}}{{Gaudi}}{2012}]{Gaudi2012}
{Gaudi} B.~S.,  2012, \mn@doi [\araa] {10.1146/annurev-astro-081811-125518},
  \href {https://ui.adsabs.harvard.edu/abs/2012ARA&A..50..411G} {50, 411}

\bibitem[\protect\citeauthoryear{{Ghezzi}, {Montet}  \& {Johnson}}{{Ghezzi}
  et~al.}{2018}]{GMJ2018}
{Ghezzi} L.,  {Montet} B.~T.,   {Johnson} J.~A.,  2018, \mn@doi [\apj]
  {10.3847/1538-4357/aac37c}, \href
  {https://ui.adsabs.harvard.edu/abs/2018ApJ...860..109G} {860, 109}

\bibitem[\protect\citeauthoryear{{Girven}, {Brinkworth}, {Farihi},
  {G{\"a}nsicke}, {Hoard}, {Marsh}  \& {Koester}}{{Girven}
  et~al.}{2012}]{Girven+2012}
{Girven} J.,  {Brinkworth} C.~S.,  {Farihi} J.,  {G{\"a}nsicke} B.~T.,  {Hoard}
  D.~W.,  {Marsh} T.~R.,   {Koester} D.,  2012, \mn@doi [\apj]
  {10.1088/0004-637X/749/2/154}, \href
  {https://ui.adsabs.harvard.edu/abs/2012ApJ...749..154G} {749, 154}

\bibitem[\protect\citeauthoryear{{Gould} et~al.,}{{Gould}
  et~al.}{2010}]{Gould+2010}
{Gould} A.,  et~al., 2010, \mn@doi [\apj] {10.1088/0004-637X/720/2/1073}, \href
  {https://ui.adsabs.harvard.edu/abs/2010ApJ...720.1073G} {720, 1073}

\bibitem[\protect\citeauthoryear{{Grishin} \& {Veras}}{{Grishin} \&
  {Veras}}{2019}]{GV2019}
{Grishin} E.,  {Veras} D.,  2019, \mn@doi [\mnras] {10.1093/mnras/stz2148},
  \href {https://ui.adsabs.harvard.edu/abs/2019MNRAS.489..168G} {489, 168}

\bibitem[\protect\citeauthoryear{{Guillochon}, {Ramirez-Ruiz}  \&
  {Lin}}{{Guillochon} et~al.}{2011}]{GRL2011}
{Guillochon} J.,  {Ramirez-Ruiz} E.,   {Lin} D.,  2011, \mn@doi [\apj]
  {10.1088/0004-637X/732/2/74}, \href
  {https://ui.adsabs.harvard.edu/abs/2011ApJ...732...74G} {732, 74}

\bibitem[\protect\citeauthoryear{{Hamers} \& {Portegies Zwart}}{{Hamers} \&
  {Portegies Zwart}}{2016}]{HPZ2016}
{Hamers} A.~S.,  {Portegies Zwart} S.~F.,  2016, \mn@doi [\mnras]
  {10.1093/mnrasl/slw134}, \href
  {https://ui.adsabs.harvard.edu/abs/2016MNRAS.462L..84H} {462, L84}

\bibitem[\protect\citeauthoryear{{Hamers}, {Antonini}, {Lithwick}, {Perets}  \&
  {Portegies Zwart}}{{Hamers} et~al.}{2017}]{Hamers+2017}
{Hamers} A.~S.,  {Antonini} F.,  {Lithwick} Y.,  {Perets} H.~B.,   {Portegies
  Zwart} S.~F.,  2017, \mn@doi [\mnras] {10.1093/mnras/stw2370}, \href
  {https://ui.adsabs.harvard.edu/abs/2017MNRAS.464..688H} {464, 688}

\bibitem[\protect\citeauthoryear{{Harding}, {Di Stefano}, {L{\'e}pine},
  {Urama}, {Pham}  \& {Baker}}{{Harding} et~al.}{2018}]{Harding+2018}
{Harding} A.~J.,  {Di Stefano} R.,  {L{\'e}pine} S.,  {Urama} J.,  {Pham} D.,
  {Baker} C.,  2018, \mn@doi [\mnras] {10.1093/mnras/stx2985}, \href
  {https://ui.adsabs.harvard.edu/abs/2018MNRAS.475...79H} {475, 79}

\bibitem[\protect\citeauthoryear{Harris et~al.,}{Harris
  et~al.}{2020}]{Harris+2020}
Harris C.~R.,  et~al., 2020, \mn@doi [Nature] {10.1038/s41586-020-2649-2}, 585,
  357

\bibitem[\protect\citeauthoryear{{Herman}, {Zhu}  \& {Wu}}{{Herman}
  et~al.}{2019}]{HZW2019}
{Herman} M.~K.,  {Zhu} W.,   {Wu} Y.,  2019, \mn@doi [\aj]
  {10.3847/1538-3881/ab1f70}, \href
  {https://ui.adsabs.harvard.edu/abs/2019AJ....157..248H} {157, 248}

\bibitem[\protect\citeauthoryear{{Hollands}, {G{\"a}nsicke}  \&
  {Koester}}{{Hollands} et~al.}{2018}]{HGK2018}
{Hollands} M.~A.,  {G{\"a}nsicke} B.~T.,   {Koester} D.,  2018, \mn@doi
  [\mnras] {10.1093/mnras/sty592}, \href
  {https://ui.adsabs.harvard.edu/abs/2018MNRAS.477...93H} {477, 93}

\bibitem[\protect\citeauthoryear{{Hunter}}{{Hunter}}{2007}]{Hunter2007}
{Hunter} J.~D.,  2007, \mn@doi [Computing in Science and Engineering]
  {10.1109/MCSE.2007.55}, \href
  {https://ui.adsabs.harvard.edu/abs/2007CSE.....9...90H} {9, 90}

\bibitem[\protect\citeauthoryear{{Johnson}, {Aller}, {Howard}  \&
  {Crepp}}{{Johnson} et~al.}{2010}]{Johnson+2010}
{Johnson} J.~A.,  {Aller} K.~M.,  {Howard} A.~W.,   {Crepp} J.~R.,  2010,
  \mn@doi [\pasp] {10.1086/655775}, \href
  {https://ui.adsabs.harvard.edu/abs/2010PASP..122..905J} {122, 905}

\bibitem[\protect\citeauthoryear{{Jones} et~al.,}{{Jones}
  et~al.}{2016}]{Jones+2016}
{Jones} M.~I.,  et~al., 2016, \mn@doi [\aap] {10.1051/0004-6361/201628067},
  \href {https://ui.adsabs.harvard.edu/abs/2016A&A...590A..38J} {590, A38}

\bibitem[\protect\citeauthoryear{{Jura}}{{Jura}}{2003}]{Jura2003}
{Jura} M.,  2003, \mn@doi [\apjl] {10.1086/374036}, \href
  {https://ui.adsabs.harvard.edu/abs/2003ApJ...584L..91J} {584, L91}

\bibitem[\protect\citeauthoryear{{Klein}, {Doyle}, {Zuckerman}, {Dufour},
  {Blouin}, {Melis}, {Weinberger}  \& {Young}}{{Klein}
  et~al.}{2021}]{Klein+2021}
{Klein} B.~L.,  {Doyle} A.~E.,  {Zuckerman} B.,  {Dufour} P.,  {Blouin} S.,
  {Melis} C.,  {Weinberger} A.~J.,   {Young} E.~D.,  2021, \mn@doi [\apj]
  {10.3847/1538-4357/abe40b}, \href
  {https://ui.adsabs.harvard.edu/abs/2021ApJ...914...61K} {914, 61}

\bibitem[\protect\citeauthoryear{{Koester}, {G{\"a}nsicke}  \&
  {Farihi}}{{Koester} et~al.}{2014}]{KGF2014}
{Koester} D.,  {G{\"a}nsicke} B.~T.,   {Farihi} J.,  2014, \mn@doi [\aap]
  {10.1051/0004-6361/201423691}, \href
  {https://ui.adsabs.harvard.edu/abs/2014A&A...566A..34K} {566, A34}

\bibitem[\protect\citeauthoryear{{Laskar}}{{Laskar}}{1997}]{Laskar1997}
{Laskar} J.,  1997, \aap, \href
  {https://ui.adsabs.harvard.edu/abs/1997A&A...317L..75L} {317, L75}

\bibitem[\protect\citeauthoryear{{Laskar}}{{Laskar}}{2008}]{Laskar2008}
{Laskar} J.,  2008, \mn@doi [\icarus] {10.1016/j.icarus.2008.02.017}, \href
  {https://ui.adsabs.harvard.edu/abs/2008Icar..196....1L} {196, 1}

\bibitem[\protect\citeauthoryear{{Laskar} \& {Petit}}{{Laskar} \&
  {Petit}}{2017}]{Laskar2017}
{Laskar} J.,  {Petit} A.~C.,  2017, \mn@doi [\aap]
  {10.1051/0004-6361/201630022}, \href
  {https://ui.adsabs.harvard.edu/abs/2017A&A...605A..72L} {605, A72}

\bibitem[\protect\citeauthoryear{{Lawson} et~al.,}{{Lawson}
  et~al.}{2012}]{Lawson+2012}
{Lawson} P.~R.,  et~al., 2012, in {Ellerbroek} B.~L.,  {Marchetti} E.,
  {V{\'e}ran} J.-P.,  eds,  Society of Photo-Optical Instrumentation Engineers
  (SPIE) Conference Series Vol. 8447, Adaptive Optics Systems III. p. 844722,
  \mn@doi{10.1117/12.925099}

\bibitem[\protect\citeauthoryear{{Li}, {Mustill}  \& {Davies}}{{Li}
  et~al.}{2022}]{Li+2022}
{Li} D.,  {Mustill} A.~J.,   {Davies} M.~B.,  2022, \mn@doi [\apj]
  {10.3847/1538-4357/ac33a8}, \href
  {https://ui.adsabs.harvard.edu/abs/2022ApJ...924...61L} {924, 61}

\bibitem[\protect\citeauthoryear{{Lithwick} \& {Wu}}{{Lithwick} \&
  {Wu}}{2011}]{LW2011}
{Lithwick} Y.,  {Wu} Y.,  2011, \mn@doi [\apj] {10.1088/0004-637X/739/1/31},
  \href {https://ui.adsabs.harvard.edu/abs/2011ApJ...739...31L} {739, 31}

\bibitem[\protect\citeauthoryear{{Lithwick} \& {Wu}}{{Lithwick} \&
  {Wu}}{2014}]{LW2014}
{Lithwick} Y.,  {Wu} Y.,  2014, \mn@doi [PNAS] {10.1073/pnas.1308261110}, \href
  {https://ui.adsabs.harvard.edu/abs/2014PNAS..11112610L} {111, 12610}

\bibitem[\protect\citeauthoryear{{Liu}, {Mu{\~n}oz}  \& {Lai}}{{Liu}
  et~al.}{2015}]{LML2015}
{Liu} B.,  {Mu{\~n}oz} D.~J.,   {Lai} D.,  2015, \mn@doi [\mnras]
  {10.1093/mnras/stu2396}, \href
  {https://ui.adsabs.harvard.edu/abs/2015MNRAS.447..747L} {447, 747}

\bibitem[\protect\citeauthoryear{{Lloyd}}{{Lloyd}}{2011}]{Lloyd2011}
{Lloyd} J.~P.,  2011, \mn@doi [\apjl] {10.1088/2041-8205/739/2/L49}, \href
  {https://ui.adsabs.harvard.edu/abs/2011ApJ...739L..49L} {739, L49}

\bibitem[\protect\citeauthoryear{{Malamud}, {Grishin}  \& {Brouwers}}{{Malamud}
  et~al.}{2021}]{MGB2021}
{Malamud} U.,  {Grishin} E.,   {Brouwers} M.,  2021, \mn@doi [\mnras]
  {10.1093/mnras/staa3940}, \href
  {https://ui.adsabs.harvard.edu/abs/2021MNRAS.501.3806M} {501, 3806}

\bibitem[\protect\citeauthoryear{{Maldonado}, {Villaver}, {Mustill}, {Chavez}
  \& {Bertone}}{{Maldonado} et~al.}{2020a}]{Maldonado+2020a}
{Maldonado} R.~F.,  {Villaver} E.,  {Mustill} A.~J.,  {Chavez} M.,   {Bertone}
  E.,  2020a, \mn@doi [\mnras] {10.1093/mnras/staa2237}, \href
  {https://ui.adsabs.harvard.edu/abs/2020MNRAS.497.4091M} {497, 4091}

\bibitem[\protect\citeauthoryear{{Maldonado}, {Villaver}, {Mustill}, {Chavez}
  \& {Bertone}}{{Maldonado} et~al.}{2020b}]{Maldonado+2020b}
{Maldonado} R.~F.,  {Villaver} E.,  {Mustill} A.~J.,  {Chavez} M.,   {Bertone}
  E.,  2020b, \mn@doi [\mnras] {10.1093/mnras/staa2946}, \href
  {https://ui.adsabs.harvard.edu/abs/2020MNRAS.499.1854M} {499, 1854}

\bibitem[\protect\citeauthoryear{{Malla} et~al.,}{{Malla}
  et~al.}{2020}]{Malla+2020}
{Malla} S.~P.,  et~al., 2020, \mn@doi [\mnras] {10.1093/mnras/staa1793}, \href
  {https://ui.adsabs.harvard.edu/abs/2020MNRAS.496.5423M} {496, 5423}

\bibitem[\protect\citeauthoryear{{Manser} et~al.,}{{Manser}
  et~al.}{2019}]{Manser+2019}
{Manser} C.~J.,  et~al., 2019, \mn@doi [Science] {10.1126/science.aat5330},
  \href {https://ui.adsabs.harvard.edu/abs/2019Sci...364...66M} {364, 66}

\bibitem[\protect\citeauthoryear{{Martin}, {Livio}, {Smallwood}  \&
  {Chen}}{{Martin} et~al.}{2020}]{Martin+2020}
{Martin} R.~G.,  {Livio} M.,  {Smallwood} J.~L.,   {Chen} C.,  2020, \mn@doi
  [\mnras] {10.1093/mnrasl/slaa030}, \href
  {https://ui.adsabs.harvard.edu/abs/2020MNRAS.494L..17M} {494, L17}

\bibitem[\protect\citeauthoryear{{Mogavero} \& {Laskar}}{{Mogavero} \&
  {Laskar}}{2021}]{ML2021}
{Mogavero} F.,  {Laskar} J.,  2021, \mn@doi [\aap]
  {10.1051/0004-6361/202141007}, \href
  {https://ui.adsabs.harvard.edu/abs/2021A&A...655A...1M} {655, A1}

\bibitem[\protect\citeauthoryear{{Mu{\~n}oz} \& {Petrovich}}{{Mu{\~n}oz} \&
  {Petrovich}}{2020}]{MP2020}
{Mu{\~n}oz} D.~J.,  {Petrovich} C.,  2020, \mn@doi [\apjl]
  {10.3847/2041-8213/abc564}, \href
  {https://ui.adsabs.harvard.edu/abs/2020ApJ...904L...3M} {904, L3}

\bibitem[\protect\citeauthoryear{{Mullally} et~al.,}{{Mullally}
  et~al.}{2021}]{MullallyJWST}
{Mullally} S.,  et~al., 2021, {1911 - A Search for the Giant Planets that Drive
  White Dwarf Accretion}, \url
  {https://www.stsci.edu/jwst/science-execution/program-information.html?id=1911}

\bibitem[\protect\citeauthoryear{{Murray} \& {Dermott}}{{Murray} \&
  {Dermott}}{1999}]{MD1999}
{Murray} C.~D.,  {Dermott} S.~F.,  1999, {Solar System Dynamics}.
Cambridge Univ. Press, Cambridge

\bibitem[\protect\citeauthoryear{{Mustill} \& {Villaver}}{{Mustill} \&
  {Villaver}}{2012}]{MV2012}
{Mustill} A.~J.,  {Villaver} E.,  2012, \mn@doi [\apj]
  {10.1088/0004-637X/761/2/121}, \href
  {https://ui.adsabs.harvard.edu/abs/2012ApJ...761..121M} {761, 121}

\bibitem[\protect\citeauthoryear{{Mustill}, {Veras}  \& {Villaver}}{{Mustill}
  et~al.}{2014}]{MVV2014}
{Mustill} A.~J.,  {Veras} D.,   {Villaver} E.,  2014, \mn@doi [\mnras]
  {10.1093/mnras/stt1973}, \href
  {https://ui.adsabs.harvard.edu/abs/2014MNRAS.437.1404M} {437, 1404}

\bibitem[\protect\citeauthoryear{{Mustill}, {Villaver}, {Veras}, {G{\"a}nsicke}
   \& {Bonsor}}{{Mustill} et~al.}{2018}]{Mustill+2018}
{Mustill} A.~J.,  {Villaver} E.,  {Veras} D.,  {G{\"a}nsicke} B.~T.,   {Bonsor}
  A.,  2018, \mn@doi [\mnras] {10.1093/mnras/sty446}, \href
  {https://ui.adsabs.harvard.edu/abs/2018MNRAS.476.3939M} {476, 3939}

\bibitem[\protect\citeauthoryear{{O'Connor} \& {Lai}}{{O'Connor} \&
  {Lai}}{2020}]{OL2020}
{O'Connor} C.~E.,  {Lai} D.,  2020, \mn@doi [\mnras] {10.1093/mnras/staa2645},
  \href {https://ui.adsabs.harvard.edu/abs/2020MNRAS.498.4005O} {498, 4005}

\bibitem[\protect\citeauthoryear{{O'Connor}, {Liu}  \& {Lai}}{{O'Connor}
  et~al.}{2021}]{OConnor+2020}
{O'Connor} C.~E.,  {Liu} B.,   {Lai} D.,  2021, \mn@doi [\mnras]
  {10.1093/mnras/staa3723}, \href
  {https://ui.adsabs.harvard.edu/abs/2021MNRAS.501..507O} {501, 507}

\bibitem[\protect\citeauthoryear{{Payne}, {Veras}, {Holman}  \&
  {G{\"a}nsicke}}{{Payne} et~al.}{2016}]{Payne+2016}
{Payne} M.~J.,  {Veras} D.,  {Holman} M.~J.,   {G{\"a}nsicke} B.~T.,  2016,
  \mn@doi [\mnras] {10.1093/mnras/stv2966}, \href
  {https://ui.adsabs.harvard.edu/abs/2016MNRAS.457..217P} {457, 217}

\bibitem[\protect\citeauthoryear{{Payne}, {Veras}, {G{\"a}nsicke}  \&
  {Holman}}{{Payne} et~al.}{2017}]{Payne+2017}
{Payne} M.~J.,  {Veras} D.,  {G{\"a}nsicke} B.~T.,   {Holman} M.~J.,  2017,
  \mn@doi [\mnras] {10.1093/mnras/stw2585}, \href
  {https://ui.adsabs.harvard.edu/abs/2017MNRAS.464.2557P} {464, 2557}

\bibitem[\protect\citeauthoryear{{Penny}, {Gaudi}, {Kerins}, {Rattenbury},
  {Mao}, {Robin}  \& {Calchi Novati}}{{Penny} et~al.}{2019}]{Penny+2019}
{Penny} M.~T.,  {Gaudi} B.~S.,  {Kerins} E.,  {Rattenbury} N.~J.,  {Mao} S.,
  {Robin} A.~C.,   {Calchi Novati} S.,  2019, \mn@doi [\apjs]
  {10.3847/1538-4365/aafb69}, \href
  {https://ui.adsabs.harvard.edu/abs/2019ApJS..241....3P} {241, 3}

\bibitem[\protect\citeauthoryear{{Petrovich} \& {Mu{\~n}oz}}{{Petrovich} \&
  {Mu{\~n}oz}}{2017}]{PM2017}
{Petrovich} C.,  {Mu{\~n}oz} D.~J.,  2017, \mn@doi [\apj]
  {10.3847/1538-4357/834/2/116}, \href
  {https://ui.adsabs.harvard.edu/abs/2017ApJ...834..116P} {834, 116}

\bibitem[\protect\citeauthoryear{{Pichierri}, {Morbidelli}  \&
  {Lai}}{{Pichierri} et~al.}{2017}]{PML2017}
{Pichierri} G.,  {Morbidelli} A.,   {Lai} D.,  2017, \mn@doi [\aap]
  {10.1051/0004-6361/201730936}, \href
  {https://ui.adsabs.harvard.edu/abs/2017A&A...605A..23P} {605, A23}

\bibitem[\protect\citeauthoryear{{Pu} \& {Lai}}{{Pu} \& {Lai}}{2018}]{PL2018}
{Pu} B.,  {Lai} D.,  2018, \mn@doi [\mnras] {10.1093/mnras/sty1098}, \href
  {https://ui.adsabs.harvard.edu/abs/2018MNRAS.478..197P} {478, 197}

\bibitem[\protect\citeauthoryear{{Pu} \& {Lai}}{{Pu} \& {Lai}}{2019}]{PL2019}
{Pu} B.,  {Lai} D.,  2019, \mn@doi [\mnras] {10.1093/mnras/stz1817}, \href
  {https://ui.adsabs.harvard.edu/abs/2019MNRAS.488.3568P} {488, 3568}

\bibitem[\protect\citeauthoryear{{Raymond} \& {Nesvorn\'{y}}}{{Raymond} \&
  {Nesvorn\'{y}}}{2020}]{RN2020}
{Raymond} S.~N.,  {Nesvorn\'{y}} D.,  2020, arXiv e-prints, \href
  {https://ui.adsabs.harvard.edu/abs/2020arXiv201207932R} {p. arXiv:2012.07932}

\bibitem[\protect\citeauthoryear{{Reffert}, {Bergmann}, {Quirrenbach},
  {Trifonov}  \& {K{\"u}nstler}}{{Reffert} et~al.}{2015}]{Reffert+2015}
{Reffert} S.,  {Bergmann} C.,  {Quirrenbach} A.,  {Trifonov} T.,
  {K{\"u}nstler} A.,  2015, \mn@doi [\aap] {10.1051/0004-6361/201322360}, \href
  {https://ui.adsabs.harvard.edu/abs/2015A&A...574A.116R} {574, A116}

\bibitem[\protect\citeauthoryear{{Ronco}, {Schreiber}, {Giuppone}, {Veras},
  {Cuadra}  \& {Guilera}}{{Ronco} et~al.}{2020}]{Ronco+2020}
{Ronco} M.~P.,  {Schreiber} M.~R.,  {Giuppone} C.~A.,  {Veras} D.,  {Cuadra}
  J.,   {Guilera} O.~M.,  2020, \mn@doi [\apjl] {10.3847/2041-8213/aba35f},
  \href {https://ui.adsabs.harvard.edu/abs/2020ApJ...898L..23R} {898, L23}

\bibitem[\protect\citeauthoryear{{Schreiber}, {G{\"a}nsicke}, {Toloza},
  {Hernandez}  \& {Lagos}}{{Schreiber} et~al.}{2019}]{Schreiber+2019}
{Schreiber} M.~R.,  {G{\"a}nsicke} B.~T.,  {Toloza} O.,  {Hernandez} M.-S.,
  {Lagos} F.,  2019, \mn@doi [\apjl] {10.3847/2041-8213/ab42e2}, \href
  {https://ui.adsabs.harvard.edu/abs/2019ApJ...887L...4S} {887, L4}

\bibitem[\protect\citeauthoryear{{Smallwood}, {Martin}, {Livio}  \&
  {Lubow}}{{Smallwood} et~al.}{2018}]{Smallwood+2018}
{Smallwood} J.~L.,  {Martin} R.~G.,  {Livio} M.,   {Lubow} S.~H.,  2018,
  \mn@doi [\mnras] {10.1093/mnras/sty1819}, \href
  {https://ui.adsabs.harvard.edu/abs/2018MNRAS.480...57S} {480, 57}

\bibitem[\protect\citeauthoryear{{Smallwood}, {Martin}, {Livio}  \&
  {Veras}}{{Smallwood} et~al.}{2021}]{Smallwood+2021}
{Smallwood} J.~L.,  {Martin} R.~G.,  {Livio} M.,   {Veras} D.,  2021, \mn@doi
  [\mnras] {10.1093/mnras/stab1077}, \href
  {https://ui.adsabs.harvard.edu/abs/2021MNRAS.504.3375S} {504, 3375}

\bibitem[\protect\citeauthoryear{{Spergel} et~al.,}{{Spergel}
  et~al.}{2015}]{Spergel+2015}
{Spergel} D.,  et~al., 2015, arXiv e-prints, \href
  {https://ui.adsabs.harvard.edu/abs/2015arXiv150303757S} {p. arXiv:1503.03757}

\bibitem[\protect\citeauthoryear{{Steckloff}, {Debes}, {Steele}, {Johnson},
  {Adams}, {Jacobson}  \& {Springmann}}{{Steckloff}
  et~al.}{2021}]{Steckloff+2021}
{Steckloff} J.~K.,  {Debes} J.,  {Steele} A.,  {Johnson} B.,  {Adams} E.~R.,
  {Jacobson} S.~A.,   {Springmann} A.,  2021, \mn@doi [\apjl]
  {10.3847/2041-8213/abfd39}, \href
  {https://ui.adsabs.harvard.edu/abs/2021ApJ...913L..31S} {913, L31}

\bibitem[\protect\citeauthoryear{{Stephan}, {Naoz}  \& {Zuckerman}}{{Stephan}
  et~al.}{2017}]{SNZ2017}
{Stephan} A.~P.,  {Naoz} S.,   {Zuckerman} B.,  2017, \mn@doi [\apjl]
  {10.3847/2041-8213/aa7cf3}, \href
  {https://ui.adsabs.harvard.edu/abs/2017ApJ...844L..16S} {844, L16}

\bibitem[\protect\citeauthoryear{{Stephan}, {Naoz}  \& {Gaudi}}{{Stephan}
  et~al.}{2021}]{Stephan+2020}
{Stephan} A.~P.,  {Naoz} S.,   {Gaudi} B.~S.,  2021, \mn@doi [\apj]
  {10.3847/1538-4357/ac22a9}, \href
  {https://ui.adsabs.harvard.edu/abs/2021ApJ...922....4S} {922, 4}

\bibitem[\protect\citeauthoryear{{Suzuki} et~al.,}{{Suzuki}
  et~al.}{2016}]{Suzuki+2016}
{Suzuki} D.,  et~al., 2016, \mn@doi [\apj] {10.3847/1538-4357/833/2/145}, \href
  {https://ui.adsabs.harvard.edu/abs/2016ApJ...833..145S} {833, 145}

\bibitem[\protect\citeauthoryear{{Teyssandier} \& {Lai}}{{Teyssandier} \&
  {Lai}}{2019}]{TL2019}
{Teyssandier} J.,  {Lai} D.,  2019, \mn@doi [\mnras] {10.1093/mnras/stz2919},
  \href {https://ui.adsabs.harvard.edu/abs/2019MNRAS.490.4353T} {490, 4353}

\bibitem[\protect\citeauthoryear{{Teyssandier} \& {Ogilvie}}{{Teyssandier} \&
  {Ogilvie}}{2016}]{TO2016}
{Teyssandier} J.,  {Ogilvie} G.~I.,  2016, \mn@doi [\mnras]
  {10.1093/mnras/stw521}, \href
  {https://ui.adsabs.harvard.edu/abs/2016MNRAS.458.3221T} {458, 3221}

\bibitem[\protect\citeauthoryear{{Teyssandier}, {Lai}  \& {Vick}}{{Teyssandier}
  et~al.}{2019}]{TLV2019}
{Teyssandier} J.,  {Lai} D.,   {Vick} M.,  2019, \mn@doi [\mnras]
  {10.1093/mnras/stz1011}, \href
  {https://ui.adsabs.harvard.edu/abs/2019MNRAS.486.2265T} {486, 2265}

\bibitem[\protect\citeauthoryear{{Touma}, {Tremaine}  \& {Kazandjian}}{{Touma}
  et~al.}{2009}]{TTK2009}
{Touma} J.~R.,  {Tremaine} S.,   {Kazandjian} M.~V.,  2009, \mn@doi [\mnras]
  {10.1111/j.1365-2966.2009.14409.x}, \href
  {https://ui.adsabs.harvard.edu/abs/2009MNRAS.394.1085T} {394, 1085}

\bibitem[\protect\citeauthoryear{{Turner} \& {Wyatt}}{{Turner} \&
  {Wyatt}}{2020}]{TW2020}
{Turner} S. G.~D.,  {Wyatt} M.~C.,  2020, \mn@doi [\mnras]
  {10.1093/mnras/stz3191}, \href
  {https://ui.adsabs.harvard.edu/abs/2020MNRAS.491.4672T} {491, 4672}

\bibitem[\protect\citeauthoryear{{Vanderbosch} et~al.,}{{Vanderbosch}
  et~al.}{2020}]{Vanderbosch+2020}
{Vanderbosch} Z.,  et~al., 2020, \mn@doi [\apj] {10.3847/1538-4357/ab9649},
  \href {https://ui.adsabs.harvard.edu/abs/2020ApJ...897..171V} {897, 171}

\bibitem[\protect\citeauthoryear{{Vanderburg} et~al.,}{{Vanderburg}
  et~al.}{2015}]{Vanderburg+2015}
{Vanderburg} A.,  et~al., 2015, \mn@doi [\nat] {10.1038/nature15527}, \href
  {https://ui.adsabs.harvard.edu/abs/2015Natur.526..546V} {526, 546}

\bibitem[\protect\citeauthoryear{{Vanderburg} et~al.,}{{Vanderburg}
  et~al.}{2020}]{Vanderburg+2020}
{Vanderburg} A.,  et~al., 2020, \mn@doi [\nat] {10.1038/s41586-020-2713-y},
  \href {https://ui.adsabs.harvard.edu/abs/2020arXiv200907282V} {585, 363}

\bibitem[\protect\citeauthoryear{{Veras}}{{Veras}}{2016}]{Veras2016}
{Veras} D.,  2016, \mn@doi [Royal Society Open Science] {10.1098/rsos.150571},
  \href {https://ui.adsabs.harvard.edu/abs/2016RSOS....350571V} {3, 150571}

\bibitem[\protect\citeauthoryear{{Veras} \& {G{\"a}nsicke}}{{Veras} \&
  {G{\"a}nsicke}}{2015}]{VG2015}
{Veras} D.,  {G{\"a}nsicke} B.~T.,  2015, \mn@doi [\mnras]
  {10.1093/mnras/stu2475}, \href
  {https://ui.adsabs.harvard.edu/abs/2015MNRAS.447.1049V} {447, 1049}

\bibitem[\protect\citeauthoryear{{Veras} \& {Scheeres}}{{Veras} \&
  {Scheeres}}{2020}]{VS2020}
{Veras} D.,  {Scheeres} D.~J.,  2020, \mn@doi [\mnras] {10.1093/mnras/stz3565},
  \href {https://ui.adsabs.harvard.edu/abs/2020MNRAS.492.2437V} {492, 2437}

\bibitem[\protect\citeauthoryear{{Veras}, {Jacobson}  \&
  {G{\"a}nsicke}}{{Veras} et~al.}{2014}]{VJG2014}
{Veras} D.,  {Jacobson} S.~A.,   {G{\"a}nsicke} B.~T.,  2014, \mn@doi [\mnras]
  {10.1093/mnras/stu1926}, \href
  {https://ui.adsabs.harvard.edu/abs/2014MNRAS.445.2794V} {445, 2794}

\bibitem[\protect\citeauthoryear{{Veras}, {Eggl}  \& {G{\"a}nsicke}}{{Veras}
  et~al.}{2015}]{VEG2015}
{Veras} D.,  {Eggl} S.,   {G{\"a}nsicke} B.~T.,  2015, \mn@doi [\mnras]
  {10.1093/mnras/stv1047}, \href
  {https://ui.adsabs.harvard.edu/abs/2015MNRAS.451.2814V} {451, 2814}

\bibitem[\protect\citeauthoryear{{Veras}, {Mustill}, {G{\"a}nsicke},
  {Redfield}, {Georgakarakos}, {Bowler}  \& {Lloyd}}{{Veras}
  et~al.}{2016}]{Veras+2016}
{Veras} D.,  {Mustill} A.~J.,  {G{\"a}nsicke} B.~T.,  {Redfield} S.,
  {Georgakarakos} N.,  {Bowler} A.~B.,   {Lloyd} M. J.~S.,  2016, \mn@doi
  [\mnras] {10.1093/mnras/stw476}, \href
  {https://ui.adsabs.harvard.edu/abs/2016MNRAS.458.3942V} {458, 3942}

\bibitem[\protect\citeauthoryear{{Veras}, {Higuchi}  \& {Ida}}{{Veras}
  et~al.}{2019}]{VHI2019}
{Veras} D.,  {Higuchi} A.,   {Ida} S.,  2019, \mn@doi [\mnras]
  {10.1093/mnras/stz421}, \href
  {https://ui.adsabs.harvard.edu/abs/2019MNRAS.485..708V} {485, 708}

\bibitem[\protect\citeauthoryear{{Villaver} \& {Livio}}{{Villaver} \&
  {Livio}}{2007}]{VL2007}
{Villaver} E.,  {Livio} M.,  2007, \mn@doi [\apj] {10.1086/516746}, \href
  {https://ui.adsabs.harvard.edu/abs/2007ApJ...661.1192V} {661, 1192}

\bibitem[\protect\citeauthoryear{{Vinson} \& {Chiang}}{{Vinson} \&
  {Chiang}}{2018}]{VC2018}
{Vinson} B.~R.,  {Chiang} E.,  2018, \mn@doi [\mnras] {10.1093/mnras/stx3091},
  \href {https://ui.adsabs.harvard.edu/abs/2018MNRAS.474.4855V} {474, 4855}

\bibitem[\protect\citeauthoryear{{Virtanen} et~al.,}{{Virtanen}
  et~al.}{2020}]{Virtanen+2020}
{Virtanen} P.,  et~al., 2020, \mn@doi [Nature Methods]
  {10.1038/s41592-019-0686-2}, \href
  {https://ui.adsabs.harvard.edu/abs/2020NatMe..17..261V} {17, 261}

\bibitem[\protect\citeauthoryear{{Volk} \& {Malhotra}}{{Volk} \&
  {Malhotra}}{2020}]{VM2020}
{Volk} K.,  {Malhotra} R.,  2020, \mn@doi [\aj] {10.3847/1538-3881/aba0b0},
  \href {https://ui.adsabs.harvard.edu/abs/2020AJ....160...98V} {160, 98}

\bibitem[\protect\citeauthoryear{{Wilson}, {Farihi}, {G{\"a}nsicke}  \&
  {Swan}}{{Wilson} et~al.}{2019}]{Wilson+2019}
{Wilson} T.~G.,  {Farihi} J.,  {G{\"a}nsicke} B.~T.,   {Swan} A.,  2019,
  \mn@doi [\mnras] {10.1093/mnras/stz1050}, \href
  {https://ui.adsabs.harvard.edu/abs/2019MNRAS.487..133W} {487, 133}

\bibitem[\protect\citeauthoryear{{Wu} \& {Lithwick}}{{Wu} \&
  {Lithwick}}{2011}]{WL2011}
{Wu} Y.,  {Lithwick} Y.,  2011, \mn@doi [\apj] {10.1088/0004-637X/735/2/109},
  \href {https://ui.adsabs.harvard.edu/abs/2011ApJ...735..109W} {735, 109}

\bibitem[\protect\citeauthoryear{{Wyatt}, {Farihi}, {Pringle}  \&
  {Bonsor}}{{Wyatt} et~al.}{2014}]{Wyatt+2014}
{Wyatt} M.~C.,  {Farihi} J.,  {Pringle} J.~E.,   {Bonsor} A.,  2014, \mn@doi
  [\mnras] {10.1093/mnras/stu183}, \href
  {https://ui.adsabs.harvard.edu/abs/2014MNRAS.439.3371W} {439, 3371}

\bibitem[\protect\citeauthoryear{{Xu}, {Dufour}, {Klein}, {Melis}, {Monson},
  {Zuckerman}, {Young}  \& {Jura}}{{Xu} et~al.}{2019}]{Xu+2019b}
{Xu} S.,  {Dufour} P.,  {Klein} B.,  {Melis} C.,  {Monson} N.~N.,  {Zuckerman}
  B.,  {Young} E.~D.,   {Jura} M.~A.,  2019, \mn@doi [\aj]
  {10.3847/1538-3881/ab4cee}, \href
  {https://ui.adsabs.harvard.edu/abs/2019AJ....158..242X} {158, 242}

\bibitem[\protect\citeauthoryear{{Xu} et~al.,}{{Xu} et~al.}{2021}]{Xu+2021}
{Xu} S.,  et~al., 2021, \mn@doi [\aj] {10.3847/1538-3881/ac2d26}, \href
  {https://ui.adsabs.harvard.edu/abs/2021AJ....162..296X} {162, 296}

\bibitem[\protect\citeauthoryear{{Zhu} \& {Wu}}{{Zhu} \& {Wu}}{2018}]{ZW2018}
{Zhu} W.,  {Wu} Y.,  2018, \mn@doi [\aj] {10.3847/1538-3881/aad22a}, \href
  {https://ui.adsabs.harvard.edu/abs/2018AJ....156...92Z} {156, 92}

\bibitem[\protect\citeauthoryear{{Zuckerman}}{{Zuckerman}}{2014}]{Zuckerman2014}
{Zuckerman} B.,  2014, \mn@doi [\apjl] {10.1088/2041-8205/791/2/L27}, \href
  {https://ui.adsabs.harvard.edu/abs/2014ApJ...791L..27Z} {791, L27}

\bibitem[\protect\citeauthoryear{{Zuckerman}, {Koester}, {Reid}  \&
  {H{\"u}nsch}}{{Zuckerman} et~al.}{2003}]{Zuckerman+2003}
{Zuckerman} B.,  {Koester} D.,  {Reid} I.~N.,   {H{\"u}nsch} M.,  2003, \mn@doi
  [\apj] {10.1086/377492}, \href
  {https://ui.adsabs.harvard.edu/abs/2003ApJ...596..477Z} {596, 477}

\bibitem[\protect\citeauthoryear{{Zuckerman}, {Koester}, {Melis}, {Hansen}  \&
  {Jura}}{{Zuckerman} et~al.}{2007}]{Zuckerman+2007}
{Zuckerman} B.,  {Koester} D.,  {Melis} C.,  {Hansen} B.~M.,   {Jura} M.,
  2007, \mn@doi [\apj] {10.1086/522223}, \href
  {https://ui.adsabs.harvard.edu/abs/2007ApJ...671..872Z} {671, 872}

\bibitem[\protect\citeauthoryear{{Zuckerman}, {Melis}, {Klein}, {Koester}  \&
  {Jura}}{{Zuckerman} et~al.}{2010}]{Zuckerman+2010}
{Zuckerman} B.,  {Melis} C.,  {Klein} B.,  {Koester} D.,   {Jura} M.,  2010,
  \mn@doi [\apj] {10.1088/0004-637X/722/1/725}, \href
  {https://ui.adsabs.harvard.edu/abs/2010ApJ...722..725Z} {722, 725}

\makeatother
\end{thebibliography}
\bibliographystyle{mnras.bst}

\label{lastpage}

\end{document}